\documentclass{aa}
\usepackage{graphicx}
\usepackage{txfonts}
\usepackage{natbib}
\usepackage{xspace}
\usepackage{color}
\usepackage{hyperref}
\graphicspath{{./figures/}{./figures/figures_appendix/}}

\newcommand{\nustar} {\textit{NuSTAR}\xspace}
\newcommand{\cepx}{Cep X\--4\,}

\newcommand{\Ecyc}{\ensuremath{E_{\mathrm{Cyc}}}}
\newcommand{\sigmacyc}{\ensuremath{\sigma_{\mathrm{cyc}}}}

\newcommand{\Efe}{\ensuremath{E_{\textrm{Fe}}}}
\newcommand{\sigmafe}{\ensuremath{\sigma_{\textrm{Fe}}}}

\def \ergcmsec{\hbox{\ensuremath{\mathrm{erg\,cm}^{-2}\,\mathrm{s}^{-1}}}}

\def \arcmin {\hbox{\ensuremath{^\prime}}}

\def \redchisq {\ensuremath{\chi^2_{\mathrm{red}}}}

\def \esplit {\ensuremath{E_{\textrm{split}}}}
\def \nbins {\ensuremath{N_{\textrm{bins}}}}

\begin{document}

\title{Energy-resolved pulse profiles of accreting pulsars:\\ Diagnostic tools for spectral features}

\author{Carlo Ferrigno\inst{1,3}
\and Antonino D'Aì\inst{2}
\and Elena Ambrosi\inst{2}
}
\institute{Department of astronomy, University of Geneva, chemin d’Écogia, 16, 1290 Versoix, Switzerland\\ \email{carlo.ferrigno@unige.ch}
\and
INAF - IASF-Palermo, via Ugo La Malfa 153, Italy
\and
INAF, Osservatorio Astronomico di Brera, Via E. Bianchi 46, I-23807, Merate, Italy
}
\date{ }


  \abstract
   {}
   {We  introduce a method for extracting spectral information from energy-resolved light curves folded at the neutron star spin period (known as pulse profiles) in accreting X-ray binaries.
   Spectra of these sources are sometimes characterized by features superimposed on a smooth continuum, such as iron emission lines and cyclotron resonant scattering features. We address here the question on how to derive \emph{quantitative} constraints on such features from energy-dependent changes in the pulse profiles.}
   {We developed a robust method for determining in each energy-selected bin the value of the pulsed fraction using the fast Fourier transform opportunely truncated at the number of harmonics needed to satisfactorily describe the actual profile. We determined the uncertainty on this value by sampling through Monte Carlo simulations a total of 1\,000 faked profiles. We rebinned the energy-resolved pulse profiles to have a constant minimum signal-to-noise ratio throughout the whole energy band. Finally we characterize the dependence of the energy-resolved pulsed fraction using a phenomenological polynomial model and search for features corresponding to spectral signatures of iron emission or cyclotron lines using Gaussian line profiles.}
   {We apply our method to a representative sample of \nustar\ observations of well-known accreting X-ray pulsars. We show that, with this method, it is possible to characterize the pulsed fraction spectra, and to  constrain the position and widths of such features with a precision comparable with the spectral  results. We also explore how harmonic decomposition, correlation, and lag spectra might be used as additional probes for detection and characterization of such features.}
   {}
   \maketitle
%

\section{Introduction} \label{sect:intro}

X-ray binary pulsars (XBPs) harbor a neutron star (NS), whose large magnetic field
($B\approx10^{10-13}$\,G) efficiently channels  accreting matter from a companion
star into the polar caps of the NS through the formation of an accretion column \citep{Basko1976}.
The conversion of accretion power to
radiation is realized through different physical processes \citep[e.g., inverse-Compton,
bremsstrahalung, synchrotron, and synchtron-Compton emission; see][and references therein]{Becker2022}.
Matter is coupled with the magnetic fields, and radiation escapes either along
the magnetic axis (pencil beam) or through the walls of the accretion column (fan beam).
As the rotational and magnetic axes in a NS are not generally aligned,
this escaping radiation would appear to a distant observer as a spin-modulated emission pattern,
similarly to what classical radio pulsars show in the radio band.

The shape of a folded pulse profile can be considered a fingerprint for the generic XBP.
This profile is determined by intrinsic characteristics of the system
as geometry, NS properties, and the mass-accretion channel.
In addition, there are well-known transient effects, such  as the instantaneous mass accretion rate,
the amount of matter  accumulated at the column base, and  local
feedback mechanisms between the magnetic field, radiation, and accreting plasma.
Owing to this complex nonlinear superposition of ingredients, added to the
severe light bending and lensing effects, the description of pulse shapes
in terms of physical models has proved to be theoretically very challenging
\citep{Nollert1989,Kraus1995,Kraus2003,Ferrigno2011a,Schonherr2014,Falkner2018}.

It has long been  known that profile shapes are energy and luminosity dependent \citep{Wang1981}.
In most cases the harmonic content tends to be larger below $\sim$10\,keV, and the pulse amplitude
generally increases with energy.
This is due to the different beaming of radiation originating from
various spectral components and/or the visibility of emitting regions for the distant observer.
An intuitive schema involves a variable contribution from pencil and fan beams.
The former is dominant at low luminosity when a radiation shock is not developed;
the latter is dominant at higher luminosity when a column is formed.
The same schema applies to radiation originating from a thermal mound plus an extended atmosphere,
which can be considered as pointing upward, and to direct radiation from the column,
which escapes mostly laterally.
Since the fan beam
can be greatly amplified by gravitational lensing when the emitting column is behind the NS,
even a limited variability can create large changes in the observed pulse for suitable geometries
\citep[see][]{Meszaros1985,Meszaros1992,Nagel1981}.

Notwithstanding this complexity,
pulse profiles still remain an invaluable tool to understand the
emission mechanism, the system geometry, and the radiation transfer in extreme conditions.
Once the pulsar ephemeris is determined, photons can be tagged in pulse phase and energy.
Then, it is straightforward to build energy-phase matrices
by binning events in appropriate channels.
These matrices are widely represented in color-coded images once pulse profiles in each energy bin are
normalized by subtracting the average and dividing by the standard deviation
\citep[e.g.,][]{Ferrigno2011,Tsygankov2006}.
To highlight quantitative properties, it is crucial to extract some
energy-dependent analytical elements that provide a more accessible and straightforward interpretation:
for instance the pulsed fraction, the harmonic content, the cross-correlation,
or the phase-lag with respect to a reference profile.
A fundamental tool for deriving most of these quantities
is the fast Fourier transform (FFT).
The FFT method can be applied to any signal shape; because it is  a simple mathematical transformation,
it requires no assumption on how it was physically generated.
The method returns a series of harmonics, where each amplitude is proportional to the power
of the signal at that frequency and the relative phase.
The first harmonic is generally called the fundamental as it carries most of the whole signal power.
The decomposition is usually truncated  when the profile is sufficiently reconstructed and, in any case,
the maximum number of harmonics cannot exceed the
\nbins\ / 2 value if \nbins\ is the number of bins in the folded pulse
profile. \citet{Alonso-Hernandez2022}  demonstrates on a  sample of XBPs sources, the energy
dependence of the amplitudes of the various harmonics derived from the spectral decomposition. A tentative
classification scheme based on the amplitude dependence of the fundamental and second harmonic is
proposed, although it could not be established if this scheme is linked to characteristic physical
properties (like spin period, accretion rate or magnetic field).

In this work we mainly focus on the pulsed fraction (PF)
properties in correlation with the energy.
The PF is a scalar value used to quantify the strength of modulated versus total emission.
However, even if the PF is always meant to refer to the strength
of the pulsed part of the total observed emission,
different operative definitions are present in the literature.
The confusion around a formal definition of the PF leads to difficulties in comparing results
obtained from different works.
Since the purpose of this work is to exploit the PF-energy dependence to derive meaningful physical constraints,
we performed a series of tests to evaluate advantages and disadvantages
in defining the best-suited operative definition of PF.
The comparison among different definitions of PF adopted in the literature and the results
of our tests are presented in Appendix \ref{app:pf_def}.

Despite the different definitions of PF, some results have emerged over  the years that relate to significant changes in the PF at energies corresponding with
some spectral features; for instance, \citet{Lutovinov2009} analyzed a sample of ten bright XRPs
observed with \emph{INTEGRAL}, showing the presence of local dips in the PF-energy plot at the spectroscopic
inferred cyclotron line energies, which they attribute to the effect of resonance absorption.
These findings have been  confirmed in many other sources,
even at lower inferred mass accretion rates \citep{Ferrigno2009, Tsygankov2007, Salganik2022, Ghising2022, Wang2022, Tobrej2023}.
Our aim here is to build a robust methodology to analyze profile changes in connection with
spectral results. This methodology will be then applied to
all available  data sets of similar sources in order to offer an additional or
alternative test for the presence
of such features in known and yet-to-be-discovered  XBPs.

This paper is organized as follows. In
Sect.~\ref{sect:method} we present the methodology through which we derive the
energy-resolved pulse profiles and the fit method applied to the PF spectrum
to constrain features in the spectrum.
In Sect.~\ref{sect:sources} we show this method applied to four different observations
from four well-known accreting XRPs. In Sect.~\ref{sect:discussion} we discuss the results and summarize our conclusions.

\section{Methods} \label{sect:method}
In this section we present  our method for deriving the energy-resolved pulse profiles and the
computation of the total PF step-by-step.
Although the method can be generally applied to any
past or present instrument,
we focus here on the specific analysis of \nustar\ \citep{Harrison2013} data sets
as this facility is, at the moment, the best-suited one to study
the behavior of pulse profiles in the hard X-ray energy
range where spectral features are often observed in XBPs.

\subsection{Data selection and reduction}
We retrieved \nustar\ observations from the public HEASARC archive.
Each observation is uniquely identified by an Observation Identification number (ObsID).
We chose four prototypical XBP sources (4U 1626-67, Her X-1, Cen X-3, and Cepheus X-4)
as detailed in the observation log in Table~\ref{tab:observing_log}.
We reprocessed the data using the \nustar\ Data Analysis
Software \textsc{nustardas} v.\,1.9.7
available in \textsc{HEASoft} v.\,6.31 along with the latest \nustar\
calibration files (CALDB v20220510). 
For each observation, we processed the data using the custom-developed \textsc{nustarpipeline}\footnote{\url{https://gitlab.astro.unige.ch/ferrigno/nustar-pipeline}} wrapper.
First, we obtained calibrated level 2 event files of the Focal Plane Module A
(FPMA) and Focal Plane Module B (FPMB).
Then, we defined the source and background circular regions encompassing 95\% of the source signal
(with a radius of about 2\arcmin) and used the \textsc{SAO ds9} software for a visual inspection.
The source region was centered on the best-known source X-ray coordinates,
whereas the background region was set
in a detector region free of contaminating sources.
None of the examined ObsIDs was affected by stray light issues.

\subsection{Filtering criteria}

\begin{figure}
    \centering
    \includegraphics[width=\columnwidth]{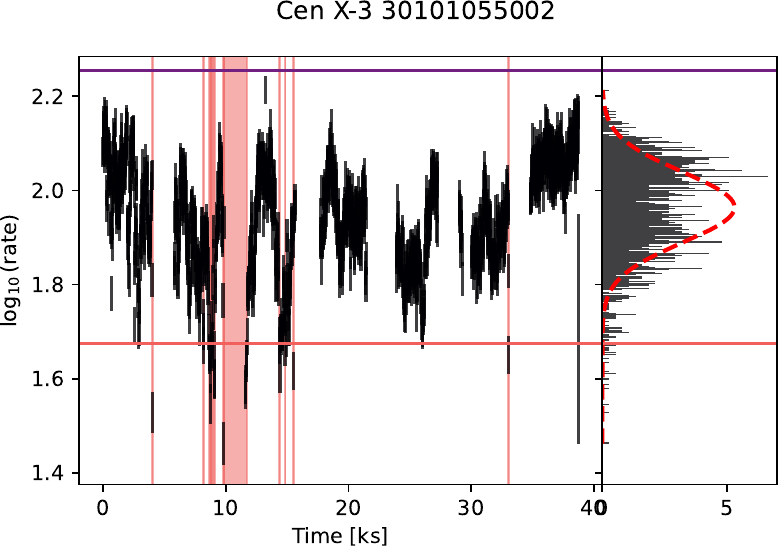}
    \caption{Filtering of Cen X-3 light curves using a log-normal distribution of counts in the 3--70\,keV band (see Sect.~\ref{sect:cen}).
        \emph{Left panel}: Light curve since the beginning of observation on 2015 November 30.
         The pink vertical bands indicate the excluded periods; the horizontal lines indicate the rate limits.
         \emph{Right panel}: Histogram of counts in logarithmic scale with horizontal lines indicating the adopted limits.
         We considered as good time intervals those encompassing the $\pm$5$\sigma$ deviations
         from the central peak value.}
    \label{fig:cenx-3gti-filter}
\end{figure}

We filtered out time intervals where the source showed
clear eclipses, dips, flares, significant spectral changes, or unusual behavior.
To this end, we extracted the source light curves in the 3--7\,keV and 7--30\,keV energy
ranges with 0.1\,s time bins, and re-binned them to ensure a minimal
signal-to-noise ratio (S/N) of 50
eventually taking the largest bin among the two.
We defined the hardness ratio for these two bands (hard/soft rates) and
studied its variation as a function of time to spot any
significant spectral change during our observation.
If no major variation is present, we studied the flux variations on the rebinned light curve
using the frequency histogram of the total 3--30\,keV rates, and excluded the intervals
falling at more than 5\,$\sigma$ from the Gaussian best-fitting peak, assuming a log-normal distribution.
We illustrate our filtering procedure in Fig.~\ref{fig:cenx-3gti-filter}.

\subsection{Pulse search, time-phase matrix, and timing analysis}

We barycentered the arrival time of each event in the Solar System frame
and  performed a Lomb-Scargle (LS) search for coherent signals.
We took the highest peak in the LS spectrum as preliminary spin frequency ($P_\mathrm{spin}$),
we checked its general consistency with the literature value, and then
we refined $P_\mathrm{spin}$ using an epoch folding search in a frequency
interval around $\pm5\%$\,$\times P_\mathrm{spin}$, after correcting for the binary
motion. When available, we used binary orbital parameters from the Fermi-GBM online
database\footnote{\url{https://gammaray.nsstc.nasa.gov/gbm/science/pulsars.html}} of XBPs.
We report these periods in Table~\ref{tab:observing_log}.

\begin{figure}
    \centering
    \includegraphics[width=\columnwidth]{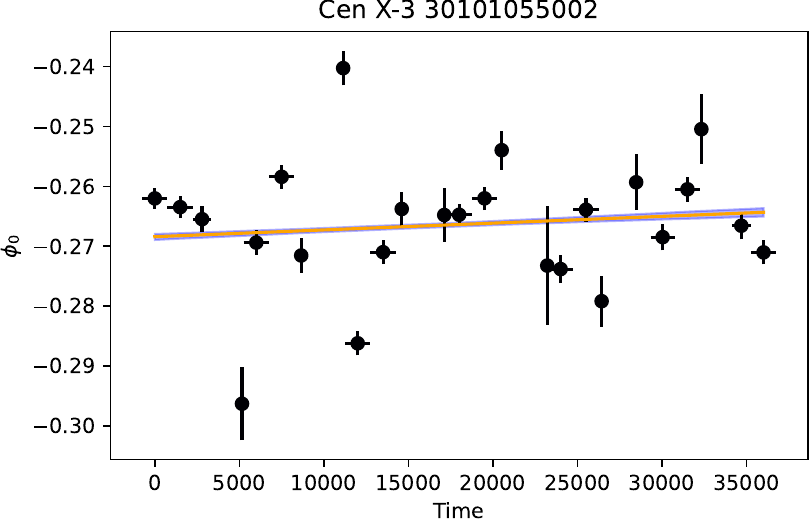}
    \caption{First-harmonic phase of time-resolved pulse profiles extracted along the Cen~X-3 observation on 2015 November 30 with time tagged since its start.
    Pulses are extracted every 1500\,s and rebinned to achieve a S/N of 15. The solid line with spread indicates the best linear trend and its 1$\sigma$ envelope. The phases are obtained by folding the light curve using the period and period derivative shown in Table~\ref{tab:observing_log}. }
    \label{fig:cenx-3timevsphaseshift}
\end{figure}

Adopting this $P_{\textrm{spin}}$,
we extracted folded pulse profiles in intervals of 1\,500\,s
for the whole duration of the observation.
We refer to this object, which contains the full set of the time-selected folded profiles, as the time-phase matrix.
We calculated each profile S/N according to the standard expression
\begin{equation}
S/N = \frac{\sum | p_i - \bar p |}{\sqrt{\sum  (\sigma_{p_i})^2}  }
,\end{equation}
where $p_i$, $\sigma_{p_i}$, and $\bar p$ are respectively the rate on the $i$ phase bin,
its uncertainty, and the average rate of the pulse profile.
We re-binned the resulting profiles to obtain a minimum S/N;
a value of $\sim$\,15 was generally satisfactory.
Subsequently, we determined the phase of
the first harmonic for each of them and plotted their values as a function of time.
We checked for the presence of a linear, or quadratic, trend in the phase values
(see Fig.~\ref{fig:cenx-3timevsphaseshift} for one of the examined sources) through a quadratic
least-squares fit of the data. Uncertainties on fit parameters were obtained by bootstraping
the time-dependent phases with 1000 realizations.
A linear trend would indicate a slightly incorrect pulse period, while a quadratic term would indicate
the presence of a significant spin-frequency derivative.
To detect such trends, we chose a threshold of 90\% significance level on the linear (quadratic)
best-fitting coefficients.
In the observation of Cen X-3 in particular, we found a significant curvature in the time dependence  of the phase,
while in all other cases the fit with a constant was always acceptable as no spin spin derivative was detectable.

\subsection{Energy-phase matrix construction and pulsed fraction}
\label{sec:enphase-pf}
When binary ephemerides are available, we   computed the photon arrival time as if they were
emitted on the binary system line of nodes. Then, we determined the spin frequency and
we accumulated energy-resolved folded profiles using a \nbins\ value of 32.
The energy bin spacing was adjusted to match the FPMA energy resolution ($\sim$\,0.4 keV at low energies,
1.2\,keV at high energies) in the whole \nustar band, resulting in 99 independent bins, in agreement with the
response matrix boundaries. Each
energy-resolved profile is saved as a row in a phase-energy matrix and stored as a file.
We extracted matrices separately for source and background and avoided a direct subtraction to preserve
Poissonian statistics. With the same reasoning, it is possible to sum the two FPM units A and B to
enhance the signal.
We adaptively rebinned the energy scale by requiring that each energy-resolved pulse of the summed
source-plus-background matrices had a
minimum S/N in each energy-dependent pulse profile.
The threshold is optimized to preserve the detection of features
in the decaying tail of the spectrum at tens of keV, and for this reason, we
started the rebinning computation from the last, highest energy, bin.

From the rebinned matrices, we compute the PF in each energy bin.
As described in Appendix~\ref{app:pf_def}, all integral RMS methods are
equivalent, at least for the S/N that we  use.
We then adopt the FFT method to  obtain   the harmonic decomposition at the same time:
\begin{equation}\label{eq:fft}
\mathrm{PF}_{\mathrm{FFT}} = \frac {\sqrt { \sum _{k = 1} ^{N_\mathrm{harm}}\mid A_{k} \mid ^2 } } {\mid A_{0}\mid }.
\end{equation}
Here $\mid A_{0}\mid$ is the average value of the pulse profile, which is the zeroth term of the
FFT transform, and the terms $A_{1...k}$ are the $k$-th terms of the
discrete Fourier transform, so that each $ \mid A_{k} \mid $ represents the amplitude of the $k$-th harmonic.

We truncated the Fourier spectral decomposition, using a number of harmonics that describe the pulse with a
statistical acceptance level of at least 10\%.
For Poissonian statistics, \citet{Kaastra2017} gave an approximate recipe for the Cstat
to obtain the expected mean and average given the data and model. They showed that under conditions
on the total number of counts that are met in our case, the Cstat  distribution is Gaussian.
In our case, the model is the Fourier decomposition at the $n$-th order. We iterated from a
minimum of $n=2$ to $n=N_{\textrm {bins}}/2$ until the Cstat
fell below its expected average plus $x$ times the expected standard deviation, where $x$ is
computed for the required significance level (e.g., 1.96 for 10\%).

We computed the uncertainty on the PF value using a bootstrap method:
for each energy-resolved profile, we simulated 1\,000 faked profiles assuming Poisson statistics in each
phase bin. We computed the average and the standard deviation of the simulated sample, verified
that the average is compatible with the value computed from data and used the
standard deviation as an estimate of the uncertainty at a 1$\sigma$ confidence level.

\subsection{Harmonic decomposition, lag, and correlation spectra} \label{sec:other_tools}

The pulse profile can be decomposed in Fourier series and, in general,
most of the power is in the first two harmonics (known as the
fundamental harmonics, or the  first harmonic and second harmonic). As these might
represent the contribution of different beam patterns \citep[e.g.,][]{Tamba2023},
it is informative to investigate
their amplitude- and phase-energy dependence separately.

\begin{figure}
    \centering
    \includegraphics[width=\columnwidth]{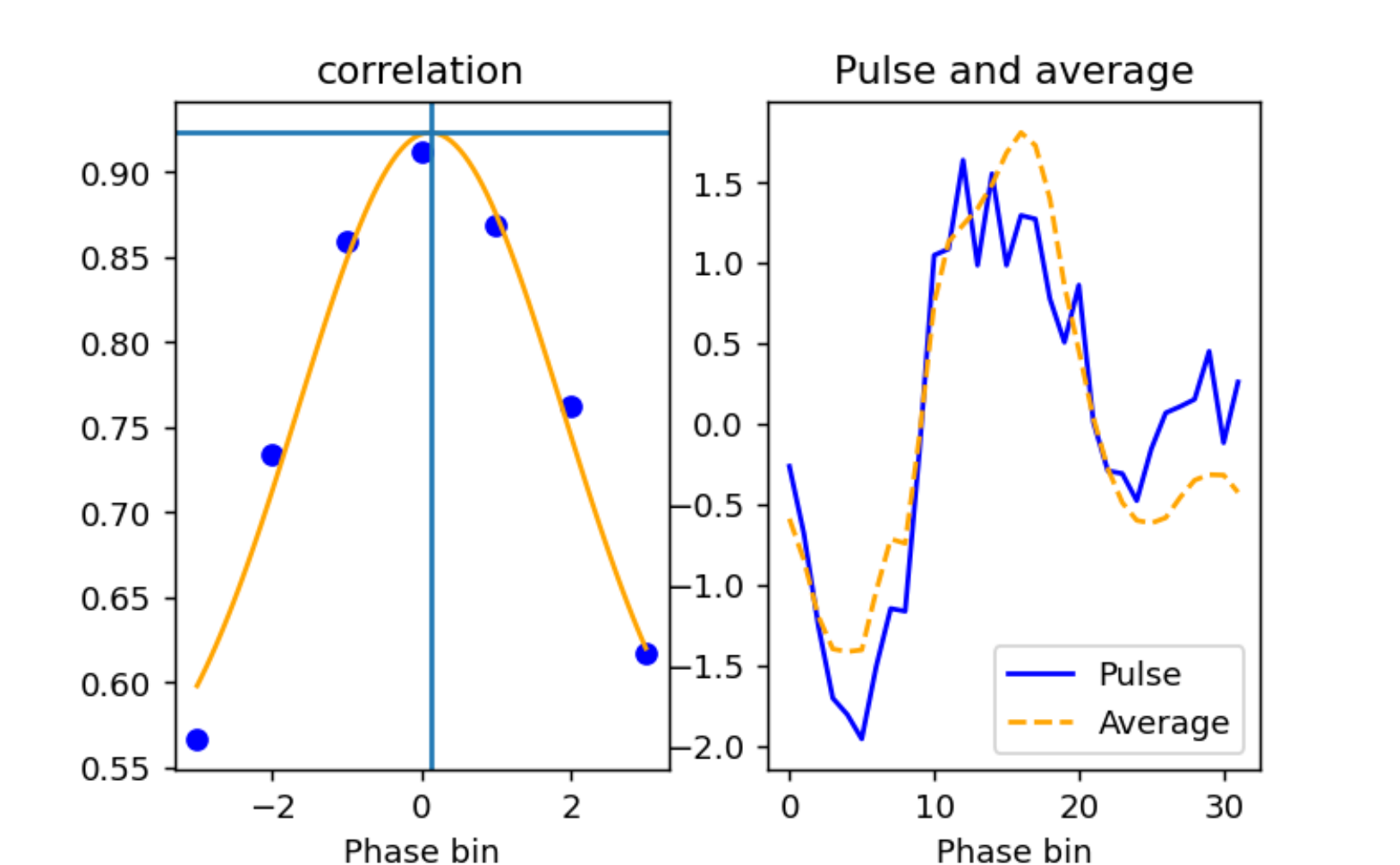}
    \caption{Example of   lag and correlation determination for the first energy bin of the Cen X-3 observation
    (see Sect.~\ref{sect:cen} for details).
    \emph{Left panel}: Points indicate correlation values as a function of the phase bin. The orange line is
    the best-fit model. The vertical solid line is drawn at the lag value, while the horizontal solid line is at the
    correlation value. \emph{Right panel}: Solid blue line is
    the pulse profile at a given energy; the dashed orange line is the average of all other energy bins.
    Both profiles are normalized to zero average and unitary standard deviation.}
    \label{fig:correlationexample}
\end{figure}

The correlation spectrum is computed taking, for each energy bin,
the correlation value between the profile at that energy bin
and the full energy-averaged profile subtracted of the contribution of the bin under exam.
We computed the discrete cross-correlation function for each energy-resolved
bin, as
\begin{equation}
c^{e}_k = \sum_n p^{e}_{n+k} \cdot \overline{p^{e}}_n\,,
\end{equation}
where $c^{e}_k$ is the correlation value for the $k$-th bin of the
pulse profile at energy $e$, $p^{e}$ is the corresponding pulse profile,
and $\overline{p}$ is the average pulse profile excluding the one at energy $e$.
Both profiles are normalized to zero average and unitary standard deviation
before computing the correlation.
However, to determine an informative lag value,
instead of taking the highest correlation value from the
discrete cross-correlation vector, we model a range of values (seven points)
around the peak of the discrete sampling using a constant plus a Gaussian function.
The corresponding best-fit Gaussian peak value is taken as the
correlation value for that energy bin, and the corresponding
peak position is used to derive its lag value in phase units
(see Fig.~\ref{fig:correlationexample} for an illustration).

We computed uncertainties by bootstrapping 100
fake energy-phase matrices using Poissonian statistics and applying
the same method to the faked data set.
Then, we took the standard deviation of
the sample as the 68\% confidence interval estimate of the relative quantity.
In all cases, we verified that the mean sample value
is compatible with the observed value.

By construction, the correlation peaks in the energy range where the statistics are the highest,
typically in the 6--10\,keV range given the hard spectrum of these sources and the
\nustar effective area.

\subsection{Modeling the pulsed fraction spectrum} \label{sect:fitting_method}
The method we used to extract and rebin the energy-dependent pulses
provides a sensible description of the energy variation of the PF.
An immediate consequence of our rebinning is the possibility to evidence
the smooth increase from low to high energies in the pulsed signal together with the
superimposed  deep features corresponding to the  $\sim 6.5$\,keV iron line (\Efe) and   the
cyclotron line (\Ecyc).
With such an accurate determination of the PF spectrum, we can model the energy dependence in
analogy with spectral fitting.

We find evidence for different trends
between the low-energy  ($\sim$\,2--15\,keV) and the high-energy ($\sim$\,15--70\,keV) bands
of the PF.
As our focus is on revealing energy-confined features,
we model the PF spectrum
in two separate energy bands defined by a splitting energy value (\esplit).
We search for a convenient \esplit\  in the range between 10 and 20\,keV, which is far from
the typical centroid energies of features.
For this purpose, we interpolate the PF  using a third-order univariate spline
from the \texttt{scipy} interpolate package \citep{scipy}.
We compute the spline derivative function and set \esplit\
equal to the energy corresponding either to a value of zero
(i.e., a flex point of the function) or to the
first derivative minimum value.


In each segment the increase in the PF with energy varies from source to source and it can be described by a
polynomial. The addition of Gaussian-like absorbing components mimics
the sudden decrease in correspondence to the Fe and cyclotron line features.
We use a Gaussian initially centered at 6.4\,keV for the iron line ($E_\mathrm{Fe}$),
with width $\sigmafe=$0.5\,keV,
while the initial values for the energy (\Ecyc) and width (\sigmacyc)
of cyclotron features are taken from spectral fitting
available in the literature.
The normalization factors $A_\mathrm{Fe}$ and $A_\mathrm{Cyc}$ are left as free parameters with
suitable initial guesses.
After allowing for a centroid-energy variation of $\pm$20\% and width swings within a few keV,
we perform a least-squares fitting.
The degree of the polynomial function is chosen adaptively for each source so that
the p-value of the best-fit model is above a certain threshold that
we set at 5\% to trade off  between model accuracy and complexity.
An additional condition, to prevent overfitting, is
that the difference of p-values obtained adding a
degree to the polynomial remains above one-fourth of the same threshold.

After performing an initial fit using the \texttt{lmfit} python package \citep{lmfit},
we explore the parameter distribution and their uncertainties
using Markov chain Monte Carlo algorithms
from the \texttt{emcee} package \citep{emcee}.
We use the Goodman \& Weare algorithm \citep{Goodman2010}
with 50 walkers, a burning phase of 500 steps, and a length between 2000 and
5000, to prevent auto-correlation effects.
From the sample, we extract the median and  68\% percentile
as the best parameter estimates. We report for each fit the reduced
chi-squared value (\redchisq) and the corresponding degrees of freedom (d.o.f.).

We automated all this process in our dedicated python package. In this way,
the same procedure can be straightforwardly applied to the amplitude of any
harmonic, being the PF their quadratic sum.

\section{Examples of the application of the method} \label{sect:sources}

We illustrate our method and the results that can be obtained
through some representative \nustar\ observations of prototypical XBP sources listed in Table~\ref{tab:observing_log}.
We used \nbins\ = 32 and employ the standard FFT RMS method to decompose the signal.
Corner plots for the posterior distribution in the parameters of the PF spectrum fit are comprehensively shown in Appendix B.

\begin{table*}
    \scriptsize
    \caption{Log of NuSTAR observations used in this work and some adopted parameters. The flux is the observed value in the 3--70\,keV band.}
    \label{tab:observing_log}
    \centering
    \begin{tabular}{lcccccccr@{}l}
        \hline
        \hline
        Source & ObsID & Start & Stop & Exposure & min S/N & Count rate\tablefootmark{a} & Flux & \multicolumn{2}{c}{$P_\mathrm{spin}$} \\
        \hline
        &        & UT & UTC & ks & & $3-70$\,keV & $10^{-9}\ergcmsec$ & \multicolumn{2}{c}{s} \\
        \hline
        4U 1626$-$67 & 30101029002 & 2015-05-04 12:26:07 & 2015-05-05 20:41:07 & 65.0 & 10 & 15.62 & 1.22 & 7.672951 & $\pm$ 0.000016 \\
        Her X-1 & 30002006005 & 2012-09-22 04:26:07 & 2012-09-22 18:36:07 & 21.9 & 16 & 96.84 & 8.22 & 1.2377185 & $\pm$ 0.0000010 \\
        Cen X-3 & 30101055002 & 2015-11-30 18:11:08 & 2015-12-01 05:01:08 & 21.4 & 16 & 63.24 & 4.35 & 4.802581 & $\pm$ 0.000003\tablefootmark{b} \\
        Cep X-4 & 80002016002 & 2014-06-18 22:01:07 & 2014-06-19 20:11:07 & 40.4 & 10 & 41.99 & 2.82 & 66.3357 & $\pm$ 0.0017 \\
        \hline
    \end{tabular}
    \tablefoot{
        \tablefoottext{a}{The FPMA count rate in the 3--70\,keV energy range.}
        \tablefoottext{b}{The period derivative is $\dot p= (3.71 \pm 0.15) \times 10^{-9}$ s\,s$^{-1}$.}
    }
\end{table*}

\subsection{4U 1626-67} \label{sect:4u1626}

4U 1626-67 is a persistent ultra-compact low-mass X-ray binary source,
whose orbital period is 42 minutes. The presence of a cyclotron line
at $\sim$\,38 keV is considered a very secure detection, confirmed multiple
times with different observatories and spectral models \citep{Orlandini1998,Coburn2002,Camero-Arranz2012}.
There is also a hint for a second cyclotron harmonic around 60 keV \citep{D'Ai2017},
but to date there has been no other independent confirmirmation. To date,
\nustar\ has observed this source  only once
for a total exposure of 65.2\,ks (in May 2015; ObsID 30101029002).

\begin{figure*}
    \centering
    \includegraphics[width=1.0\linewidth]{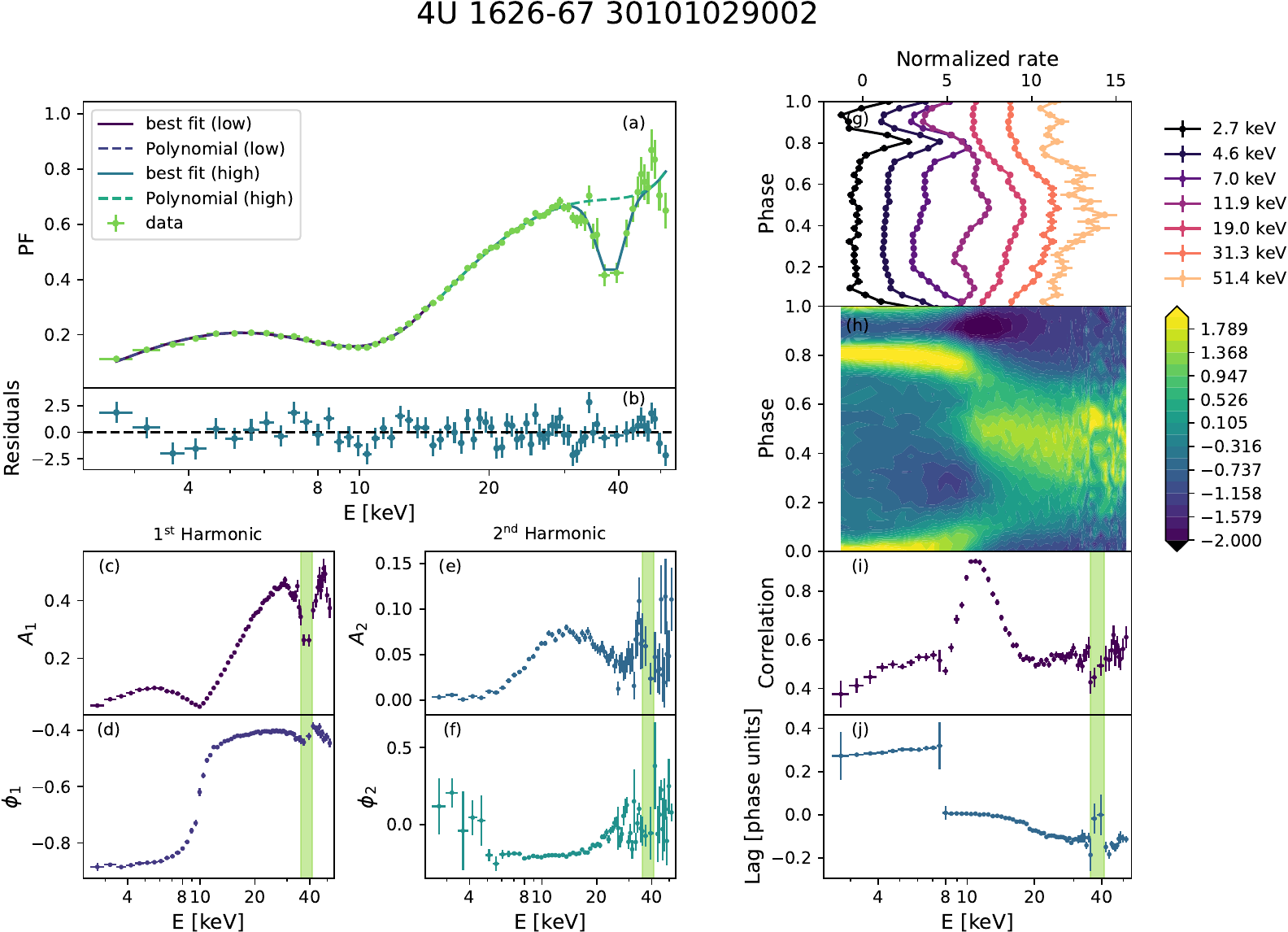}
    \caption{Pulse profile main properties for 4U 1626$-$67 in ObsID 30101029002.
        \emph{Panel (a)}: pulsed fraction (green points) and its best-fit model (solid lines); the
        polynomial functions are also shown.
        \emph{Panel (b)}: Fit residuals.
        \emph{Panels (c--f)}: Phases and amplitudes of the first ($A_1$, $\phi_1$) and second ( $A_2$, $\phi_2$) harmonics.
        The vertical colored bands indicate the energy and width of the Gaussian functions fitted to the pulsed fraction.
        \emph{Panel (g)}: Selection of normalized pulse profiles at equally logarithmic spaced energies, horizontally shifted for clarity.
        In each bin  the pulse was normalized by subtracting the average and dividing by the standard deviation.
        \emph{Panel (h)}: color-map representation of the normalized pulse profiles as a function of energy. The thin lines represent 20 equally-spaced contours.
        \emph{Panel (i)}: Cross-correlation between the pulse profile in each energy band and the average profile.
        \emph{Panel (j)}: Corresponding phase lag. The colored vertical bands are the same as in  panels (d--f).
            }
    \label{fig:4U_1626-6730101029002_summary_plot}
\end{figure*}

The PF spectrum in this source shows a significant dynamical range,
from $\sim$\,0.1 at a few keV up to a value close to 1 for the last energy bins.
The PF generally increases with energy, but
between 5 and 10\,keV there is a clear trend reversal,
so that the continuum appears to be composed of two broad humps.
From a visual inspection (Fig.~\ref{fig:4U_1626-6730101029002_summary_plot} panel (a)),
we find a local moderately broad feature
at the cyclotron line energy, but no clear residual pattern in the iron K$\alpha$ range.

\begin{figure}
    \centering
    \includegraphics[width=\columnwidth]{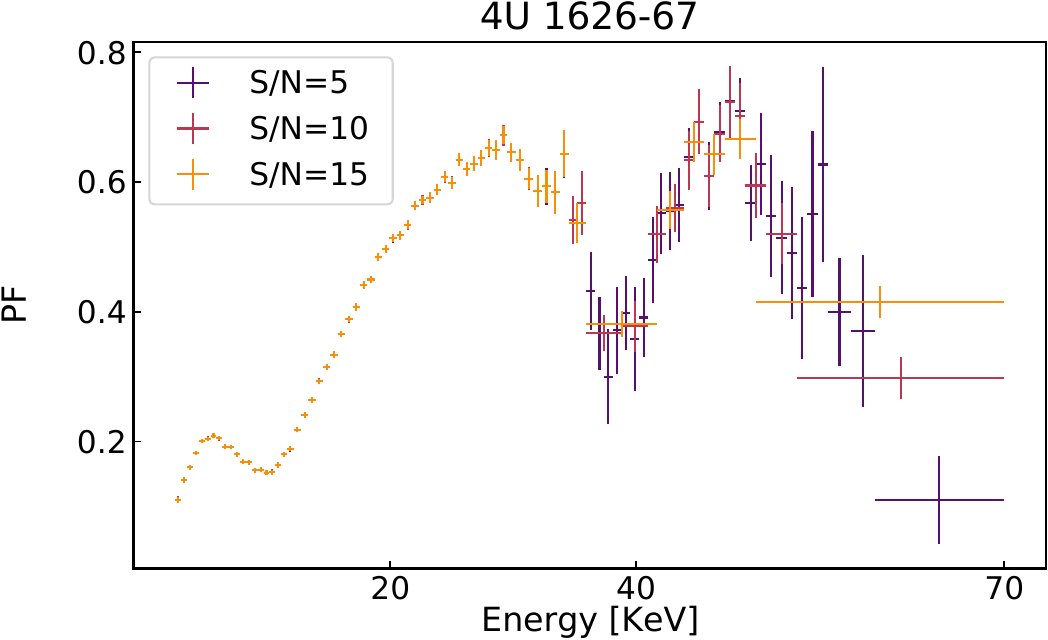}
    \caption{PF$_{\mathrm{FFT}}$ obtained with different S/N values for the pulse profile of 4U 1626 $\--$ 67.}
    \label{fig:4u1626pf}
\end{figure}

To best describe the feature at the cyclotron line energy,
we should choose an appropriate S/N for the phase-energy matrix.
In Fig.~\ref{fig:4u1626pf} we show the PF spectrum for three different choices of S/N (5, 10, and 15,
resulting in 85, 71, and 63 independent bins, respectively). Because of the much higher statistics at low
energies, this rebinning affects only the high-energy part of the PF spectrum.\footnote{This investigation is made for all sources, but we report it only for this one for illustrative purposes.}
There is a good indication that the PF spectrum sharply decreases above 50 keV. Since in this work we are mostly interested in deriving spectral
constraints on the cyclotron feature, we looked for a sufficient number of bins to describe
the spectral shape without being forced to describe in detail the PF continuum in all the
available bands. For this reason, we chose the PF spectrum with a minimum S/N\,=\,10 (which is
a good compromise between number of bins and statistics at the cyclotron line), but we removed the
last data point in the fitting of the data, as its
sharp PF drop would lead to an  undesirable higher
complexity in the continuum polynomial description.

Given this complex continuum for this source,
the fit of the PF was done separately for the soft (2--13 keV)
and hard (13--53 keV) X-ray bands.
The soft band is well fitted using a fourth-order polynomial (\redchisq\ =\,1.1, for 17 d.o.f.),
and no notable local residual is apparent.
For the hard band a third-order polynomial is required
to achieve a satisfactory fit (\redchisq\ =\,1.4, for 40 d.o.f.)
together with a moderately broad absorption Gaussian close to the cyclotron energy.
Panels (a) and (b) of Fig. \ref{fig:4U_1626-6730101029002_summary_plot}
respectively show the data, together with the best-fitting model, and
residuals in units of sigma.
We show in Table \ref{tab:4u1626_cyc}
the best-fitting parameters and the line parameters in direct comparison with the spectral fit result from \citep{D'Ai2017}.

\begin{table}
    \centering
    \caption{Best-fit parameters of the 4U 1626$-$67 ObsID 30101029002 pulse and amplitude models compared to spectral results.
        The last column gives the values
        from the spectral best fit of the same data set \citep[see Model \textsc{BWM3} of Table 1 in][]{D'Ai2017}.}
    \label{tab:4u1626_cyc}
    \begin{tabular}{lr@{}lr@{}lr@{}lr@{}ll}
        \hline
        \hline
         & \multicolumn{2}{c}{PF} & \multicolumn{2}{c}{$1^\mathrm{st}$} & \multicolumn{2}{c}{Spectral} &  \\
        \hline
        $\chi^2_\mathrm{red,lo}$/d.o.f. &  1.1 &/17 &   1.6 &/13 & -- & -- & \\
        $\chi^2_{\mathrm{red},hi}$/d.o.f. &  1.4 &/40 &   1.1 &/40 & -- & -- & \\
        $n_\mathrm{pol}^\mathrm{(lo)}$ &4 & & 6 & & -- & -- & \\
        $n_\mathrm{pol}^\mathrm{(hi)}$ &3 & & 5 & & -- & -- & \\
        $E_\mathrm{split}$ &13.21 & & 12.33 & & -- & -- & keV\\
        $A_\mathrm{Cyc}$ &-2.16 &$\pm$0.14 & -2.2 &$_{-0.3}^{+0.2}$ & 23.0 &$\pm$0.9 & \\
        $E_\mathrm{Cyc}$ &38.29 &$\pm$0.15 & 39.07 &$\pm$0.19 & 37.90 &$\pm$0.15 & keV\\
        $\sigma_\mathrm{Cyc}$ &3.06 &$\pm$0.20 & 3.6 &$\pm$0.2 & 6.0 &$\pm$0.3 & keV\\
         \hline
    \end{tabular}
    \tablefoot{
        $\chi^2_{\mathrm{red,lo}}$ and $\chi^2_{\mathrm{red},hi}$ are the reduced $\chi^2$ for the lower-
        and higher-energy sections, respectively,
        with $n_\mathrm{pol}^\mathrm{(lo)}$ and $n_\mathrm{pol}^\mathrm{(hi)}$
        the corresponding polynomial orders.
        $E_\mathrm{split}$ is the energy at which we separate the two regions.
}
\end{table}

The energy cyclotron value and the width between our fitting
PF spectrum method and the spectral fit appear to be reasonably
close in agreement. The statistical discrepancy, which is a few standard deviations,
does not take into account the systematic uncertainty in the
spectral estimations from the application of possible different continua as the spectral fit errors
are only the statistical uncertainties from a singular adopted continuum model fit.
For instance, the cyclotron line width is 3.9$^{+0.3}_{-0.9}$ keV if the spectral adopted continuum
is a power law with a high-energy cutoff \citep{D'Ai2017},
which is fully compatible with the estimate derived from the PF spectrum fit.

There are marginal differences in parameter estimation if  the same best fit is considered for the
PF spectrum with a lower or higher S/N (e.g., the energy position is 38.0\,$\pm$\,0.4\,keV
and 38.6$\pm$0.4\,keV, the line width is 3.9$\pm$0.5 and 3.3$\pm$0.4\,keV for
the PF spectrum at S/N\,=\,5 and S/N\,=\,15,  respectively).

As shown in panels (c) and (e) of Fig. \ref{fig:4U_1626-6730101029002_summary_plot},
the feature in the PF spectrum at the
cyclotron energy is clearly resolved only for the amplitude of the fundamental.
Applying the same model used for the PF spectrum, similar values for the
cyclotron line energy and width are retrieved (see Table~\ref{tab:4u1626_cyc}),
whereas the second harmonic shows a significant lower amplitude and a noisy scatter of data points,
which does not permit any firm conclusion; we do not report any fit result in this case.

The correlation spectrum (panel (h) of Fig.~\ref{fig:4U_1626-6730101029002_summary_plot}) peaks
around 10\,keV with rapid decrease  in the low-energy and the high-energy wings.
There is a hint for a small decrease in the correlation value
with respect to the adjacent bins at the cyclotron line energy.
Similarly, for the same energy bins, in the lag spectrum
(panel (i) of Fig.~\ref{fig:4U_1626-6730101029002_summary_plot})
there is some evidence for an increase in the lag with respect to the adjacent bins.
It is clear from the color representation in panel (g) that the
pulse profile has a strong shape discontinuity, which explains these characteristics.

\subsection{Her X-1} \label{sect:her}
Her X-1 is a well-known XBP located at 5.0$_{-0.7} ^{+0.6} $ kpc \citep{Arnason2021},
where a 1.4 M$_{\odot}$ NS accretes from a low- to intermediate-mass donor star
(HZ Her, 1.6 - 2 M$_{\odot} $ ) through Roche-lobe overflow.
The NS spins with a period P$_{\mathrm{spin}}$ = 1.2 s \citep{Giacconi1971}.
The orbital period is 1.7\,d \citep{Staubert2007} and also shows   superorbital
modulation of about 35\,d
due to the precession of its accretion disk (\citealt{Brumback2021}, \citealt{Scott2000}, and references therein).

\nustar\ has observed Her~X-1 on multiple occasions.
Here we analyze one of the earliest observations performed on 2012 September 22
(ObsID 30002006005) when the count rate  was high
and stable along the entire observation \citep[see][for further details]{Furst2013}.
It is known that this prototypical cyclotron line source shows a secular and luminosity-dependent shift in the
energy line position \citep{Staubert2020}.
As the benchmark for this work, we take as spectral reference the average value (\Ecyc\ =\,37.5\,$\pm$\,0.5\,keV)
obtained from the the use of different best-fitting continuum models
for this particular ObsID, as found in \citet{Furst2013};
similarly, we take the average value from the best-fitting models
for the cyclotron width  (\sigmacyc\ =\,6.9 $\pm$ 1.4 keV).
The spectrum of this observation also shows a complex emission pattern of broad iron lines,
where the most significant one has a position \Efe\,=\,6.55 keV, and a broadness \sigmafe\,=\,0.82 keV.
The presence of a second cyclotron harmonic at energies around 70 keV has been
sporadically reported \citep{Enoto2008}, but there is
no spectral evidence for its presence in this particular data set,
given also the limited energy coverage of the FPMs at such energies.

\begin{figure*}
    \centering
    \includegraphics[width=1.0\linewidth]{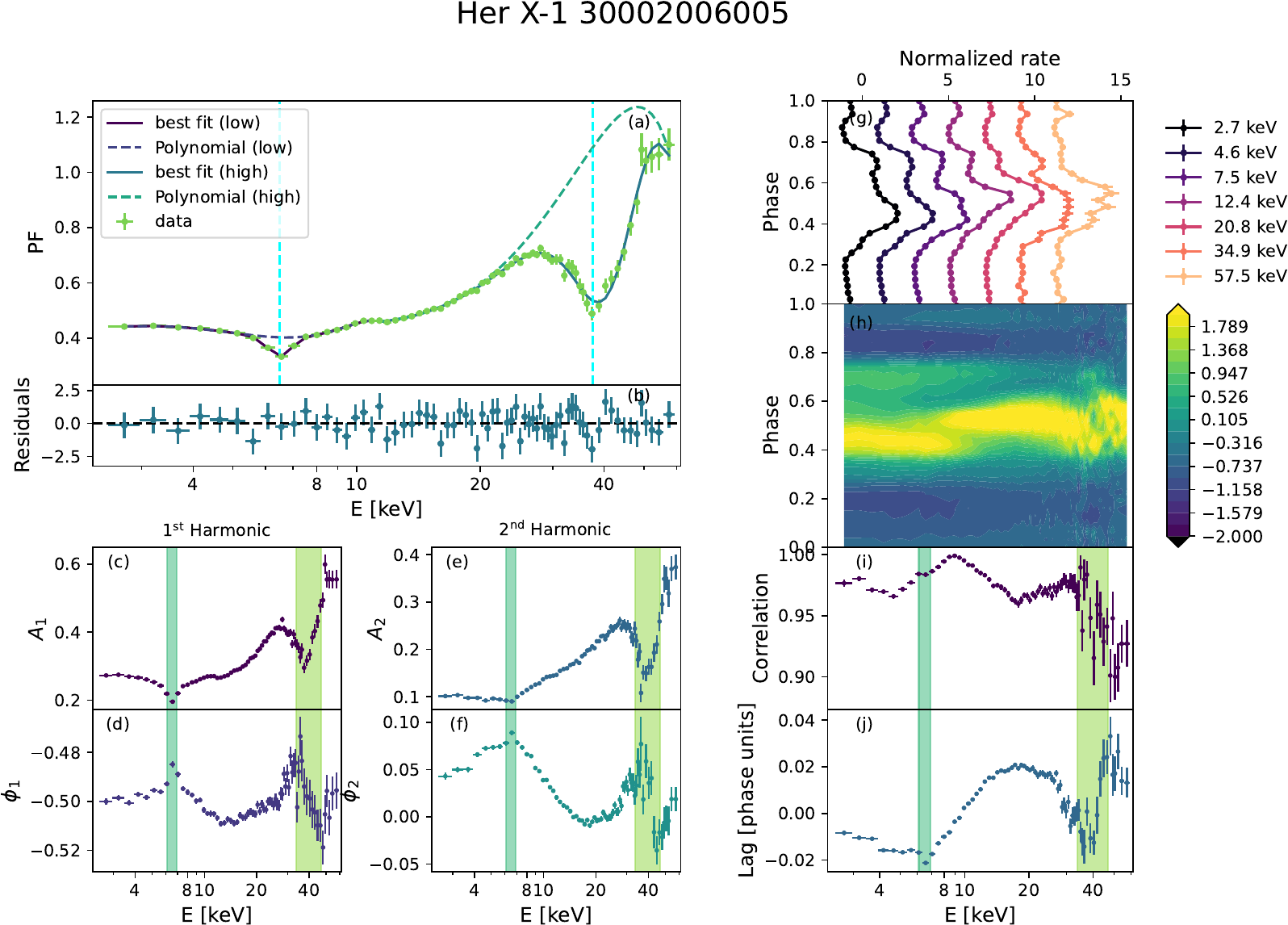}
    \caption{Pulse profile main properties for Her X-1 in ObsID 30002006005.
        \emph{Panel (a)}: pulsed fraction (green points) and its best-fit model (solid lines);
        polynomial functions are also shown.
        \emph{Panel (b)}: Fit residuals.
        \emph{Panels (c--f)}: Phases and amplitudes of the first ($A_1$, $\phi_1$) and second ( $A_2$, $\phi_2$) harmonics.
        The vertical colored bands indicate the energy and width of the Gaussian functions fitted to the pulsed fraction.
        \emph{Panel (g)}: Selection of normalized pulse profiles at equally logarithmic spaced energies, horizontally shifted for clarity.
        In each bin the pulse was normalized by subtracting the average and dividing by the standard deviation.
        \emph{Panel (h)}: Color-map representation of the normalized pulse profiles as a function of energy. The thin lines represent 20 equally-spaced contours.
        \emph{Panel (i)}: Cross-correlation between the pulse profile in each energy band and the average profile.
        \emph{Panel (j)}: Corresponding phase lag. The colored vertical bands are the same as in panels (d--f).
            }
    \label{fig:Her_X-130002006005_summary_plot}
\end{figure*}

We set the minimum S/N to 16 for the energy-phase matrix, extracted with \nbins =\,32,
obtaining 74 independent bins.
The PF spectrum is split in two separate fits at 11.2\,keV. The PF shows a
general increasing trend with energy, with two clear Gaussian-like drops
at the iron and cyclotron line energies (Fig.~\ref{fig:Her_X-130002006005_summary_plot} panel (a))

For the low-energy band a fourth-degree polynomial
describes the
data very well (\redchisq\ =\,0.7,  10 d.o.f.), together with an absorption Gaussian at energy $\Efe=6.50\pm0.02$\,keV and $\sigmafe=0.42\pm0.02$\,keV.
The presence of this feature is statistically very
significant (\redchisq\ =\,49 for 13 d.o.f., without the line).

For the high-energy band, a third-degree polynomial is needed (\redchisq\ =\,1.0,  47 d.o.f.)
for an acceptable fit of the continuum. The feature corresponding to the drop associated
with the cyclotron absorption appears smooth and well described using a Gaussian profile.
The Gaussian position is 40.44\,$\pm$\,0.15 keV and \sigmacyc\ is 6.82\,$\pm$\,0.19 keV.
 The cyclotron line position and the width are both roughly consistent with the spectral values,
but the feature in the PF spectrum is at slightly  higher energy than the corresponding spectral results.
In correspondence with the spectral cyclotron position
energy ($\sim$\,37 keV, highlighted by the dotted vertical cyan line in panel (a)
of Fig.~\ref{fig:Her_X-130002006005_summary_plot}),
we note the residuals hinting at the presence of a narrower core, which
is within the general structure of the line. This suggests that the PF
energy dependence might be more complex than a single structure.

\begin{figure*}
    \centering
   \begin{tabular}{cc}
\includegraphics[width=\columnwidth]{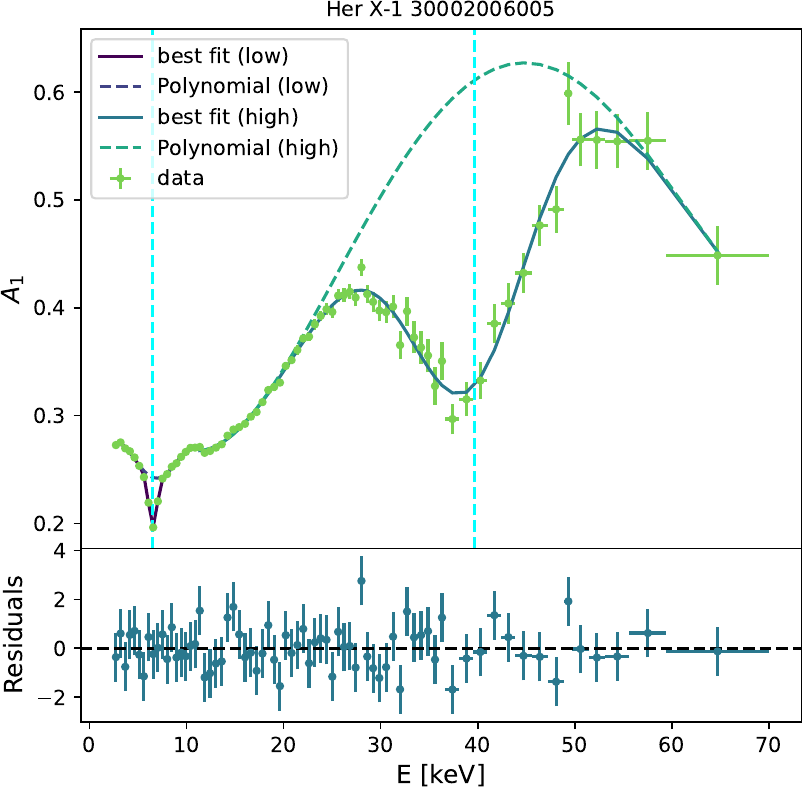} &
\includegraphics[width=\columnwidth]{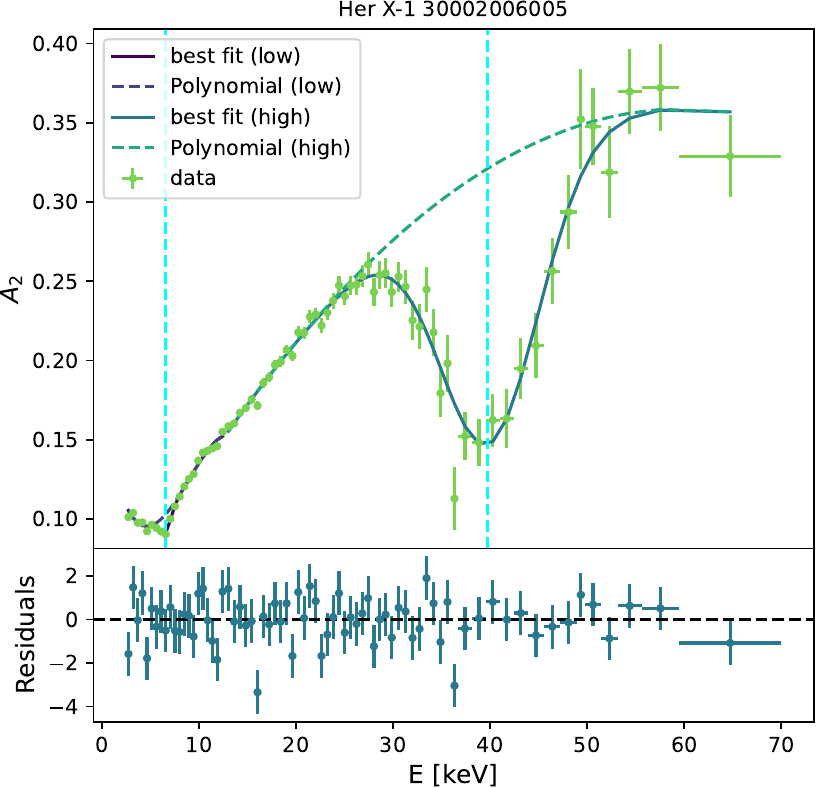}  \\
    \end{tabular}
    \caption{Her X-1 ObsID 30002006005. Shown are data, model, and best-fitting residuals for the fundamental and
    the second harmonic amplitudes. The dashed blue vertical lines are the initial line fit locations, set at 6.4\,keV and 39\,keV.}
    \label{fig:herx1_harm}
\end{figure*}

The energy-resolved amplitudes of the first two harmonics closely resemble
the PF spectrum (panels (c) and (e) of Fig.~\ref{fig:Her_X-130002006005_summary_plot}).
It is possible in both cases to apply
the same model and retrieve the line parameters, which
remain  consistent with each other (see Table~\ref{tab:herx1_harmonics} and Fig.~\ref{fig:herx1_harm}).
In particular, the presence of the iron line feature is much more clearly seen
in the fundamental rather than in the second harmonic.
We note  significant phase drifts at the iron and cyclotron line energies (panels (d) and (f) of Fig.~\ref{fig:Her_X-130002006005_summary_plot}). While there are no notable features in the correlation spectrum,
the lag spectrum shows clear local drops in the feature energies (both for
the iron and cyclotron lines) confirming that the full pulse is affected (panels (i) and (j) of Fig.~\ref{fig:Her_X-130002006005_summary_plot}).

\begin{table*}
    \caption{Her X-1 ObsID 30002006005. Comparison of best-fit parameters for the fundamental amplitude,
    second harmonic amplitude, and for the PF computed by averaging over all
    the fitted harmonic amplitudes. Spectral fit results from \citet{Furst2013} using a
    power law with a high-energy cutoff.}
    \label{tab:herx1_harmonics}
    \centering
    \begin{tabular}{lr@{}lr@{}lr@{}lr@{}ll}
        \hline
        \hline
        & \multicolumn{2}{c}{PF} & \multicolumn{2}{c}{$1^\mathrm{st}$} & \multicolumn{2}{c}{$2^\mathrm{nd}$} & \multicolumn{2}{c}{Spectral} &  \\
        \hline
        $\chi^2_\mathrm{red,lo}$/d.o.f. &  0.7 &/10 &   0.5 &/10 &   1.5 &/14 & -- & -- & \\
        $\chi^2_{\mathrm{red},hi}$/d.o.f. &  1.0 &/47 &   1.2 &/47 &   1.2 &/45 & -- & -- & \\
        $n_\mathrm{pol}^\mathrm{(lo)}$ &4 & & 4 & & 3 & & -- & -- & \\
        $n_\mathrm{pol}^\mathrm{(hi)}$ &3 & & 3 & & 2 & & -- & -- & \\
        $E_\mathrm{split}$ &11.17 & & 11.01 & & 12.82 & & -- & -- & keV \\
        $A_\mathrm{Fe}$ &-0.072 &$\pm$0.002 & -0.0486 &$\pm$0.0016 & -0.019 &$_{-0.004}^{+0.003}$ & (6.3 &$\pm$0.7) $\times10^{-3}$ & \\
        $E_\mathrm{Fe}$ &6.503 &$\pm$0.008 & 6.524 &$\pm$0.008 & 6.51 &$\pm$0.05 & 6.55 &$\pm$0.05 & keV\\
        $\sigma_\mathrm{Fe}$ &0.418 &$\pm$0.010 & 0.422 &$\pm$0.010 & 0.64 &$_{-0.08}^{+0.11}$ & 0.82 &$_{-0.10}^{+0.13}$ & keV\\
        $A_\mathrm{Cyc}$ &-10.4 &$\pm$0.7 & -6.9 &$\pm$0.5 & -2.36 &$\pm$0.09 & 0.6 &$\pm$0.3 & \\
        $E_\mathrm{Cyc}$ &40.44 &$\pm$0.15 & 40.32 &$\pm$0.16 & 40.10 &$\pm$0.16 & 37.4 &$\pm$0.2 & keV\\
        $\sigma_\mathrm{Cyc}$ &6.82 &$\pm$0.19 & 7.2 &$\pm$0.2 & 5.18 &$\pm$0.14 & 5.8 &$\pm$0.3 & keV\\
        \hline
    \end{tabular}
    \tablefoot{
        $\chi^2_{\mathrm{red,lo}}$ and $\chi^2_{\mathrm{red},hi}$ are the reduced $\chi^2$ for the lower-
        and higher-energy sections, respectively,
        with $n_\mathrm{pol}^\mathrm{(lo)}$ and $n_\mathrm{pol}^\mathrm{(hi)}$
        the corresponding polynomial orders.
        $E_\mathrm{split}$ is the energy at which we separate the two regions.
    }
\end{table*}

\subsection{Cen X-3} \label{sect:cen}
Cen~X-3 was the first X-ray pulsar to be recognized as such \citep{Giacconi1971}.
Since then, it has been observed
with most, if not all, X-ray facilities, becoming a benchmark for theory and observations.
The Cen~X-3 system is located at $6.4^{+1.0}_{-1.4}$\,kpc \citep{Arnason2021} and
comprises a O6–8 III donor star, V779 Cen, with a mass of
20.5$\pm$0.7\,M$_\odot$ \citep{vanderMeer2007} and a radius of 12\,R$_\odot$ orbiting in 2.08\,d
with a neutron star of mass 1.34$^{+0.16}_{-0.14}$\,M$\odot$ spinning at a period of 4.08\,s.
The X-ray light curve shows eclipses, spanning  20\% of the orbital period,
due to the high inclination angle \citep[about 70$^{\circ}$;][]{Ash1999}.

The accretion flow comes from the donor wind
and it is believed to be mediated by an accretion disk \citep[e.g.,][]{Suchy2008}.
The phase-averaged spectrum of the source can be described by an absorbed Comptonization spectrum
with several features.
We note a complex of fluorescence lines due to iron \citep[][]{Ebisawa1996, Iaria2005}
and a cyclotron resonant scattering feature at $\sim$30\,keV \citep{Nagase1992, Santangelo1998}.
The overall power law with high-energy cutoff shape of the continuum
needs some adjustments like the introduction of a bump around 10\,keV \citep{Suchy2008}
or partial covering \citep{Farinelli2016,Tomar2021}.
The pulse profile was decomposed assuming identical beam patter
by two poles by \citet{Kraus1996} who find a nonantipodal configuration.
This idea has been successfully exploited to interpret the IXPE
X-ray polarization results \citep{Tsygankov2022}.
\nustar\ observed Cen X-3 from 2015 November 30 to December 1,
with an elapsed time of 38.7\,ks (ObsID: 30101055002).
Different analysis find prominent iron emission around 6.4\,keV and cyclotron absorption at about 29\,keV,
regardless of the specific model or data selection \citep{Tomar2021,Thalhammer2021,Tamba2023}.

We determined the spin period after making orbital correction based on the ephemeris by \citet{Finger2010}
at 4.80265(2)\,s, and a spin period derivative $\dot p= \left(3.71 \pm 0.15\right) \times 10^{-9}$\,s\,s$^{-1}$,
compatible with existing determinations.
We extracted the energy phase matrix with 32 \nbins\ and a minimum S/N\,=\,16.
We determined the energy dependent PF and note that typically we can describe
the pulse profile using from two to eight harmonics.
We truncate our analysis just above 40\,keV, as the last energy bin is rather wide and noisy.

\begin{figure*}
    \centering
    \includegraphics[width=1.0\linewidth]{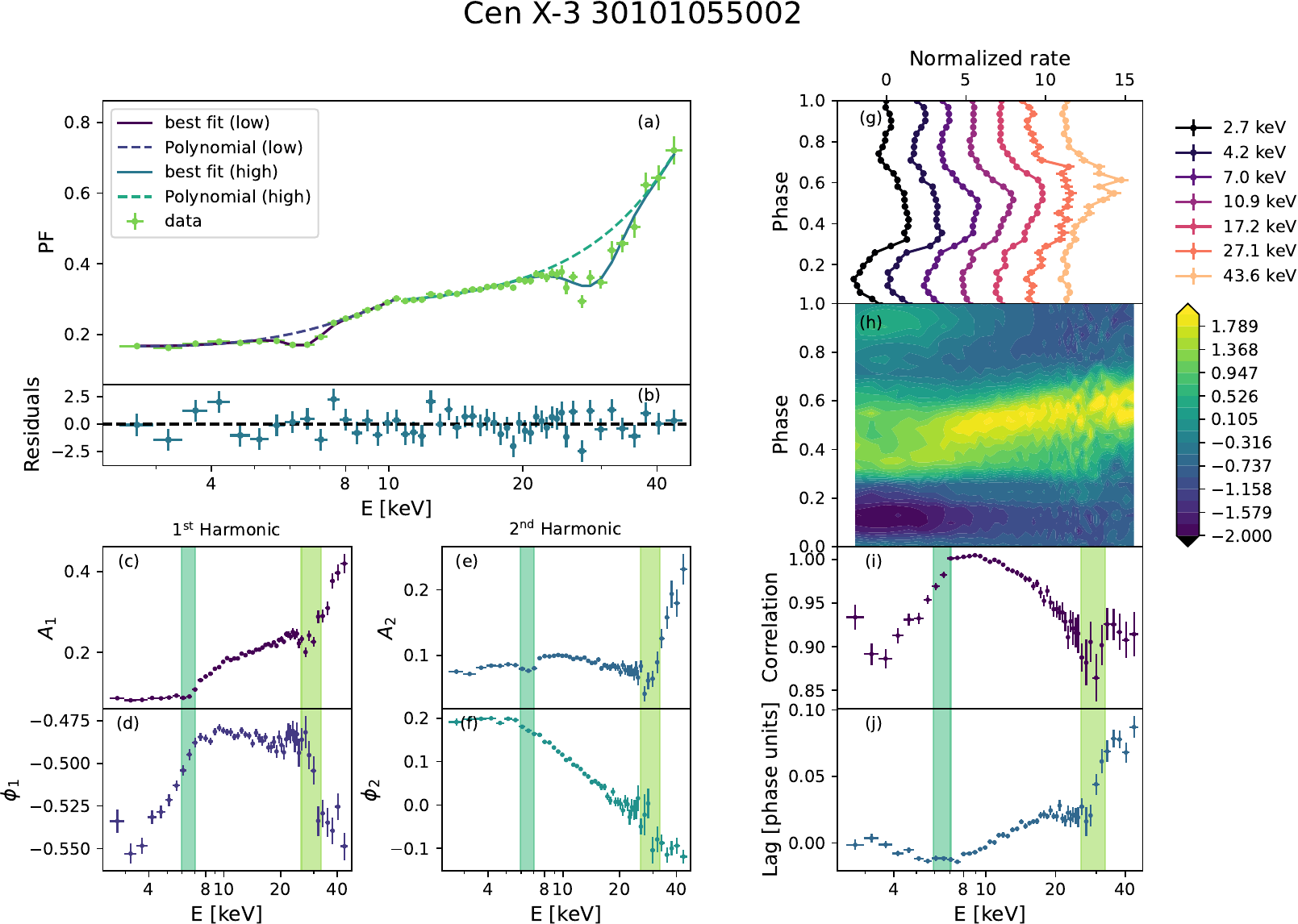}
    \caption{Pulse profile main properties for Cen X-3.
        \emph{Panel (a)}: pulsed fraction (green points) and its best-fit model (solid lines);
        polynomial functions are also shown.
        \emph{Panel (b)}: Fit residuals.
        \emph{Panels (c--f)}: Phases and amplitudes of the first ($A_1$, $\phi_1$) and second ( $A_2$, $\phi_2$) harmonics.
        The vertical colored bands indicate the energy and width of the Gaussian functions fitted to the pulsed fraction.
        \emph{Panel (g)}: Selection of normalized pulse profiles at equally logarithmic spaced energies, horizontally shifted for clarity.
        In each bin the pulse was  normalized  by subtracting the average and dividing by the standard deviation.
        \emph{Panel (h)}: Color-map representation of the normalized pulse profiles as function of energy. The thin lines  represent 20 equally-spaced contours.
        \emph{Panel (i)}: Cross-correlation between the pulse profile in each energy band and the average profile.
        \emph{Panel (j)}: Corresponding phase lag. The colored vertical bands are the same  as in panels (d--f).
    }
    \label{fig:cenx-330101055002summaryplot}
\end{figure*}

As we show in
Fig.~\ref{fig:cenx-330101055002summaryplot} panels (a--b), there is a change in slope between 10 and 20 keV: our
algorithm finds a zero derivative at 10.9\,keV. Moreover,
we see two very pronounced features modeled as
Gaussians centered at $6.44\pm0.02$ and $29.8\pm0.8$\,keV
in addition to a relatively smooth continuum, described by second-order polynomial functions.
This is enough to provide a p-value of 1\% in the fit.
The amplitude of the first and second harmonics (panels c and e in Fig.~\ref{fig:cenx-330101055002summaryplot})
can be roughly described by the same kind of model. However, the
suppression in correspondence of the cyclotron line is not significant for the
first harmonic, while the second harmonic shows a much wider feature, suggesting that
most of the cyclotron signal is encoded in a structure repeating twice in the pulse profile.
The comparison of best-fit parameters for the PF, the first two harmonics, and the
literature spectral results are reported in Table~\ref{tab:cen_x-3_line_parameters},
showing a solid consistency of the feature centroid energies, while the width
of the PF feature is slightly smaller than that in the spectrum.
The corner plots in Appendix~\ref{app:emcee_plots} confirm the goodness of the PF description.
In Fig.~\ref{fig:cenx-330101055002summaryplot} panels (d) and (f),
we note that the phases of the first harmonic presents some wiggles at the cyclotron energy,
a change in trend around 6--7\,keV.

\begin{table*}
    \caption{Best-fit parameters of the Cen X-3 ObsID 30101055002 pulse and amplitude models compared to spectral results \citep{Tomar2021}.}
    \label{tab:cen_x-3_line_parameters}
    \centering
    \begin{tabular}{lr@{}lr@{}lr@{}lr@{}ll}
        \hline
        \hline
        & \multicolumn{2}{c}{PF} & \multicolumn{2}{c}{$1^\mathrm{st}$} & \multicolumn{2}{c}{$2^\mathrm{nd}$} & \multicolumn{2}{c}{Spectral} &  \\
        \hline
         & \multicolumn{2}{c}{PF} & \multicolumn{2}{c}{$1^\mathrm{st}$} & \multicolumn{2}{c}{$2^\mathrm{nd}$} & \multicolumn{2}{c}{Spectral} &  \\
        \hline
        $\chi^2_\mathrm{red,lo}$/d.o.f. &  2.0 &/10 &   2.0 &/8 &   0.8 &/17 & -- & -- & \\
        $\chi^2_\mathrm{red,hi}$/d.o.f. &  1.1 &/29 &   1.6 &/27 &   0.8 &/23 & -- & -- & \\
        $n_\mathrm{pol}^\mathrm{(lo)}$ &3 & & 2 & & 2 & & -- & -- & \\
        $n_\mathrm{pol}^\mathrm{(hi)}$ &2 & & 7 & & 2 & & -- & -- & \\
        $E_\mathrm{split}$ &10.71 & & 9.40 & & 13.82 & & -- & -- & keV\\
        $A_\mathrm{Fe}$ &-0.060 &$\pm$0.003 & -0.0241 &$\pm$0.0018 & -0.0265 &$\pm$0.0016 & 0.117 &$_{-0.011}^{+0.007}$ & \\
        $E_\mathrm{Fe}$ &6.485 &$\pm$0.018 & 6.43 &$\pm$0.02 & 6.55 &$\pm$0.03 & 6.67\tablefootmark{a} & & keV\\
        $\sigma_\mathrm{Fe}$ &0.56 &$\pm$0.02 & 0.47 &$\pm$0.03 & 0.57 &$\pm$0.03 & 0.53\tablefootmark{a} &$\pm$0.03 & keV\\
        $A_\mathrm{Cyc}$ &-1.14 &$\pm$0.13 & 0.06 &$_{-0.15}^{+0.11}$ & -0.35 &$_{-0.06}^{+0.05}$ & 0.40 &$\pm$0.13 & \\
        $E_\mathrm{Cyc}$ &29.5 &$\pm$0.3 & 27 &$\pm$3 & 29.1 &$\pm$0.3 & 30.3 &$\pm$0.6 & keV\\
        $\sigma_\mathrm{Cyc}$ &3.8 &$\pm$0.3 & 3.8 &$_{-1.2}^{+0.9}$ & 3.1 &$_{-0.4}^{+0.5}$ & 5.0 &$\pm$0.9 & keV\\
        \hline
    \end{tabular}
\tablefoot{
         $\chi^2_{\mathrm{red,lo}}$ and $\chi^2_{\mathrm{red},hi}$ are the reduced $\chi^2$ for the lower-
         and higher-energy sections, respectively,
         with $n_\mathrm{pol}^\mathrm{(lo)}$ and $n_\mathrm{pol}^\mathrm{(hi)}$
         the corresponding polynomial orders.
        $E_\mathrm{split}$ is the energy at which we separate the two regions.\\
    \tablefoottext{a}{\citet{Tomar2021} model the iron line with a complex of three lines at 6.4, 6.67, and 6.97\,keV.
        Here we report the width of the central one, which is fitted to the data and representative of the full width. }
}
\end{table*}

To asses the coherence of the pulse profiles along the energy range, we computed the cross-correlation
and the lag from average. As shown in Fig.~\ref{fig:cenx-330101055002summaryplot} panels (h--i),
there is a strong dependence of the pulse profile on energy, while the lags start to increase
again above the cyclotron energy after showing a plateau between 20 and 30\,keV.
This is different from the behavior of the phases of the main Fourier components,
but the reason is not clear.
To have an illustration of the pulse profile energy dependence, we show them as a color map in
panel (g) of Fig.~\ref{fig:cenx-330101055002summaryplot},
where we can see the gradual phase drift and the
change in the pulse shape.

\subsection{Cepheus X-4} \label{sect:cep}

Cepheus X-4 (hereafter \cepx) is an accreting X-ray pulsar with spin period of 66.2\,s and
a Be star as a donor \citep{Ulmer1973} at a distance of 3.8\,$\pm$\,0.6 kpc \citep{Bonnet-Bidaud1998}.
\nustar\ observed it twice during a bright outburst that occurred in 2014.
Here we  examine the first observation (ObsID 80002016002),
which was performed close to the peak of the outburst.
\citet{Furst2015} analyzed the broadband spectrum, finding a two-component continuum:
a soft black-body emission of temperature $\sim$\,0.9 keV and a power-law emission with
a high-energy Fermi-Dirac cutoff. Overimposed on this continuum they found evidence of an iron fluorescence
line at energy 6.5\,keV, and  a cyclotron absorption line at 30.4$\pm$0.2\,keV with a width of 5.8$\pm$0.4\,keV.
Residuals in the red part of the wing of the cyclotron line required the addition of another Gaussian
absorption component at energies of 19.0$\pm$0.5\,keV and $\sigma$=2.5$\pm$0.4\,keV.
\citet{Vybornov2017} revisited this spectral study, expanding the analysis
to include pulse-resolved spectra. This study substantially confirmed the energy position of the
iron fluorescence line, the position of the 30 keV cyclotron line, but claimed in addition the detection
of a possible harmonic at 54.8\,$\pm$\,0.5 keV. Differently from the first study, these authors
chose to model the residuals in the red wing of the line using an absorption line at $\sim$\,10 keV.
We limit ourselves to the feature at 30\,keV.

\begin{figure*}
    \centering
    \includegraphics[width=1.0\linewidth]{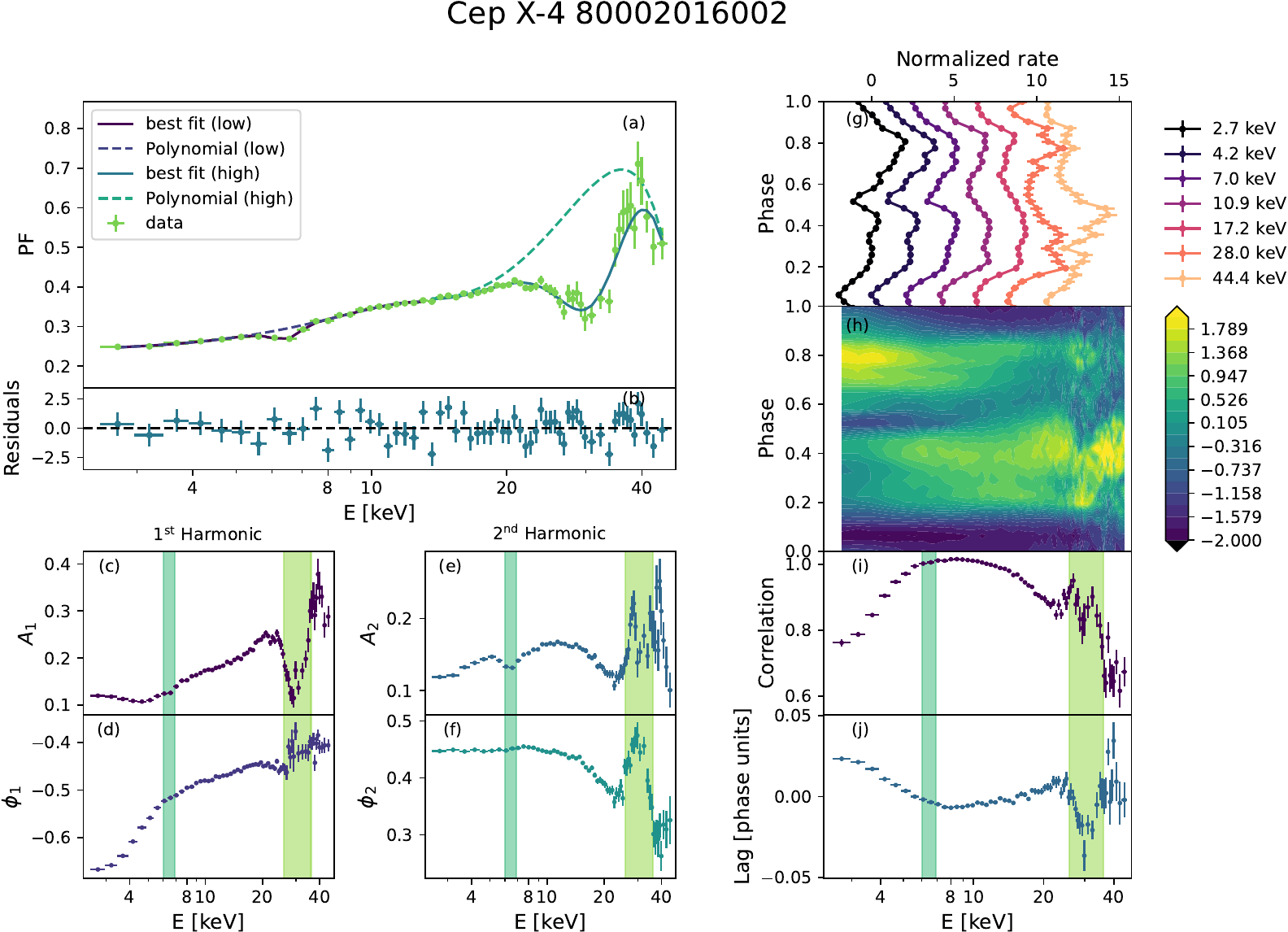}
    \caption{Pulse profile main properties for Cep X-4.
        \emph{Panel (a)}: pulsed fraction (green points) and its best-fit model (solid lines);
        polynomial functions are also shown.
        \emph{Panel (b)}: Fit residuals.
        \emph{Panels (c--f)}: Phases and amplitudes of the first ($A_1$, $\phi_1$) and second ( $A_2$, $\phi_2$) harmonics.
        The vertical colored bands indicate the energy and width of the Gaussian functions fitted to the pulsed fraction.
        \emph{Panel (g)}: Selection of normalized pulse profiles at equally logarithmic spaced energies, horizontally shifted for clarity.
        In each bin the pulse was normalized by subtracting the average and dividing by the standard deviation.
        \emph{Panel (h)}: Color-map representation of the normalized pulse profiles as function of energy. The thin lines represent 20 equally-spaced contours.
        \emph{Panel (i)}: Cross-correlation between the pulse profile in each energy band and the average profile.
        \emph{Panel (j)}: Corresponding phase lag. The colored vertical bands the same  as in panels (d--f).
    }
    \label{fig:Cep_X-480002016002_summary_plot}
\end{figure*}

After the determination of the spin period, $P_{\mathrm{spin}} = 66.33568(5)$ s, we computed the energy phase matrix
for different \nbins\ and minimum S/N and the corresponding spectrum of the PF,
choosing a value of 10 to favor the high-energy energy resolution and excluded the last energy bin,
which was too noisy.
As seen in Fig. \ref{fig:Cep_X-480002016002_summary_plot} panels (a)--(b),
the PF increases smoothly from 3 to 20\,keV, starting from 0.25 up to a maximum of
about 0.4 with a clear drop at about 6.5\,keV.
The PF trend changes dramatically beyond 20\,keV, with a well-defined and deep decrease at \Ecyc\,
followed by a sharp increase up to values 0.6--0.7 and above.

\begin{table*}
    \caption{Best-fit parameters of the Cep X-4 ObsID 80002016002 pulse and amplitude models compared to spectral results \citep{Furst2015}.}
    \label{tab:cep_x-4_line_parameters}
    \centering
    \begin{tabular}{lr@{}lr@{}lr@{}ll}
        \hline
        \hline
        & \multicolumn{2}{c}{PF} & \multicolumn{2}{c}{$1^\mathrm{st}$} & \multicolumn{2}{c}{Spectral} &  \\
        \hline
        $\chi^2_\mathrm{red,lo}$/d.o.f. &  1.4 &/15 &   0.8 &/8 & -- & -- & \\
        $\chi^2_\mathrm{red,hi}$/d.o.f. &  1.5 &/35 &   1.3 &/39 & -- & -- & \\
        $n_\mathrm{pol}^\mathrm{(lo)}$ &3 & & 6 & & -- & -- & \\
        $n_\mathrm{pol}^\mathrm{(hi)}$ &3 & & 3 & & -- & -- & \\
        $E_\mathrm{split}$ &13.35 & & 11.18 & & -- & -- & keV\\
        $A_\mathrm{Fe}$ &-0.0308 &$\pm$0.0019 & (-8 &$\pm$2) $\times10^{-3}$ & (1.39 &$\pm$0.16) $\times10^{-3}$ & \\
        $E_\mathrm{Fe}$ &6.444 &$\pm$0.019 & 6.62 &$\pm$0.04 & 6.47 &$\pm$0.03 & keV\\
        $\sigma_\mathrm{Fe}$ &0.46 &$\pm$0.02 & 0.33 &$\pm$0.05 & 0.42 &$\pm$0.05 & keV\\
        $A_\mathrm{Cyc}$ &-4.1 &$\pm$0.7 & -2.08 &$\pm$0.12 & 20 &$_{-4}^{+5}$ & \\
        $E_\mathrm{Cyc}$ &31.00 &$\pm$0.18 & 30.32 &$\pm$0.10 & 30.39 &$_{-0.14}^{+0.17}$ & keV\\
        $\sigma_\mathrm{Cyc}$ &5.3 &$\pm$0.3 & 3.75 &$\pm$0.12 & 5.8 &$\pm$0.4 & keV\\
        \hline
    \end{tabular}
    \tablefoot{
        $\chi^2_{\mathrm{red,lo}}$ and $\chi^2_{\mathrm{red},hi}$ are the reduced $\chi^2$ for  the lower-
        and higher-energy sections, respectively,
        with $n_\mathrm{pol}^\mathrm{(lo)}$ and $n_\mathrm{pol}^\mathrm{(hi)}$
        the corresponding polynomial orders.
        $E_\mathrm{split}$ is the energy at which we separate the two regions.
    }
\end{table*}

As reported in Table~\ref{tab:cep_x-4_line_parameters}, a polynomial of third degree
is necessary to obtain a p-value greater than 0.05 in the PF model in both energy bands,
at the splitting energy \esplit =\,13.35\,keV.
The best-fitting value for \Ecyc\ is 31.00$\pm$0.18\,keV, which is $\approx 1$\,keV higher than the values
reported by \citet{Furst2015} for the same observation, while the cyclotron width is a close
match. The position and width of the iron line are consistent with the spectral results.

In panels (c) and (e) of Fig.~\ref{fig:Cep_X-480002016002_summary_plot}, we show the energy dependence
of the first and second harmonic amplitudes.
The first harmonic clearly shows a drop at \Ecyc, even if its relative position appears lower with respect
to the PF spectrum, while there is a significant difference in the width.
The feature in the second harmonic amplitude looks narrow and bound between two apparent peaks,
the higher of which has low significance.
Given the relatively low signal, our modeling with a polynomial and Gaussian introduces a degeneracy
between the functions, preventing us from performing a meaningful parameter determination.
Interestingly,
the behavior of the phases for the second harmonic seems to describe a broader shape
analogous to the PF.
The iron line is clearly evident in the second harmonic rather than in the fundamental harmonic.
The lag spectrum reveals a drop at \Ecyc,\ which appears very similar to the drop observed in the PF spectrum,
whereas the correlation
values show a more structured complex, with a sharp discontinuity in the right wing of the cyclotron line,
reflecting a change in shape in the pulse.

\section{Discussion} \label{sect:discussion}

We explored various methods for reducing, encoding, and visualizing the information
contained in the energy- and phase-dependent pulsed emission of classical X-ray binary pulsars.
The aim of this work is  to illustrate the general methodology,
where we have mainly focused on one particular aspect, the ways that we  can  exploit the energy-dependent pulse information,
particularly the PF spectrum, to investigate the presence and characteristics of features
in the energy spectrum. As noted in Sect.\ref{sect:intro}, it has long been
known that the pulse profile changes in connection to some characteristic spectral features,
such as the cyclotron line, where it was easy to note clear drops in the
PF quantity, when represented in appropriate energy bins.
We considered if, and to what extent, it is possible to evolve
from a qualitative assessment on the generic scatter plot of PF points versus energy,
to physical quantitative estimates  of the main parameters responsible
for such  pulse changes. We analyzed different operative
definitions of PF and chose one that was able to best capture the fast
variations  of the pulse shape at the line energy (see details of  this work in   Appendix A)
and we finely tuned the construction of the energy-phase matrix. A base matrix is originally
created according to the instrument energy resolution, but as the typical spectra
of XBP show exponential decay, the high-energy part is mostly  very noisy. We found that an opportune rebinning, keeping the S/N above
a certain threshold, eliminates the intrinsic statistical noise, still preserving the
general aspect of the PF spectrum. There is a marginal dependence of the
PF spectrum on the number of phase bins and the minimum S/N, so that the matrix construction
has, at the end, a few user-defined parameters. However, the choice of the final setup
does not affect substantially the parameter determination,
as long as the user has a focus on the description
of the PF changes at the line energies. We clearly show this via some
examples (see Fig.~\ref{fig:4u1626pf} and Fig.~\ref{fig:her_snr}).

We approached the description of the PF spectrum
in purely phenomenological terms.
We used a polynomial function, keeping its degree free to vary
according to the PF complexity of each source or observation  to model data points;
the advantages of this choice are that this continuum does not
make any assumptions on the physics of a particular source; its terms
are uncorrelated orthogonal functions; and  it is easy and fast to add higher terms, which
keeps a good control of the statistical significance for the addition of each new term.
We note that, to our knowledge, there is still no physical procedure to predetermine how
the PF behaves in correlation with energy, even for a source
whose physical characteristics are already well determined.

Such a simple continuum is, however, unable to fully capture the PF behavior along
the whole \nustar\ energy band for PF spectra dominated by high statistics,
if we require a reasonably low polynomial degree, as
the values
of data points are so well determined that the function has to closely pass through them
to achieve an acceptable fit. There is a subtle modeling problem: any drop in the PF spectrum
could be described with a continuum polynomial, if its degree were sufficiently high.
However,
it is a scientist's choice to distinguish the part of the continuum and the locus
of line-like features to be modeled differently. The same issue affects spectral analysis in which cyclotron features can sometimes be model-dependent, and their detection significance varies
according to the chosen continuum.
For this work, we show our method starting from a small sample of sources,
where the spectral features (both iron and cyclotron lines) have been assessed
multiple times by different authors and instruments, in order to make these features robust
benchmarks of our method.
Because we  knew the exact positions of the features,
we found it appropriate to split the data fitting into a low-energy (generally 2--15 keV) and a high-energy (15--70 keV) band.
At the moment, the purpose of this division is to reduce the complexity
of the fit and better constrain local parameters,
but we note that for the case of the double-humped PF continuum
of 4U\,1626$-$67 (panel (g), Fig. \ref{fig:4U_1626-6730101029002_summary_plot}),
this might suggest a change in the observed emitting regions
or in the superposition of different spectral components
peaking in different energy bands \citep[see also][]{Tsygankov2021}.

By examining this small sample of XBPs, we note that local PF drops in the
spectrum are present at the inferred cyclotron and iron line energies
in all cases. These drops are well described with Gaussian profiles.
This is a first important point
that has emerged from our study. Spectral fits of cyclotron lines use either Gaussian
or Lorentzian profiles; it is believed that lines might also hide much more complexity
as features are formed not in a single slab, but likely along a geometrical extended
environment with different physical conditions \citep[e.g.,][]{Schonherr2007}, the superposition
of which might skew the profile or make it much more structured. In the case of the
examined PF spectra, we only found a hint of a possible more complex structure of the PF
at the cyclotron line energy for Her X-1 (Sect. \ref{sect:her}).

The Fourier spectral decomposition allowed us to test the
presence and strength of the features, also for single harmonics. In this work we examined the
contributions to the total PF from the fundamental and second harmonic, by taking their
energy-dependent amplitude variations. In all cases we find, as expected, that the
fundamental amplitude closely reproduces the drop in the PF spectrum at the same energies,
whereas for the second harmonic amplitude, for weaker sources like 4U~1626$-$67 and Cep~X~4,
there is a noisy scatter that prevents us from deriving meaningful constraints.
By directly comparing the PF best-fit model parameters with the values obtained from spectral modeling
already present in the literature, two aspects appear remarkable to us: the general consistency
of the line positions and widths of the features, with a scatter of a few percent from
the corresponding spectral parameters values, and the relative error on the determination
of such quantities, which is comparable to the spectral uncertainties.
This makes the direct modeling of PF spectra a viable and very sensitive probe
of the same spectral features.

In the low-energy band, all four examined sources show emission lines from neutral
or ionized Fe K$\alpha$ lines in the energy spectrum, though not all at the same
strength. These emission lines are most likely produced in the pulsar external magnetosphere,
either in a truncated accretion disk or by the hot wind of the companion star, or from
the accretion stream between the pulsar and the companion. In any of these regions,
these fluorescence photons are not clocked with the pulse spin, so that it is expected
that their contribution to the PF spectrum mimics an absorption feature, whose amplitude
and position must be closely linked to the spectral feature parameters.
The very broad and luminous iron line in the Her~X-1 system, is indeed
prominent in the PF spectrum. The line position
closely matches the position of the spectral line peak, whereas the line width appears smaller
by a factor of $\sim$\,2. This is most likely due to the very complex pattern of lines,
not just a single one, that is present in this source \citep{Kosec2022}.
The \nustar\ spectral resolution
does not permit a clear spectroscopic investigations, and disentangling the
line contributions over the continuum might easily lead to over- or underestimating
the physical parameters in the spectral fit.

For the cases of Cep~X-4 and Cen~X-3 iron lines, the PF
corresponding lines match very well with the spectral results.
For 4U~1626$-$67, the PF spectrum is likely not of sufficient statistical quality to detect this feature;
in fact, the line might be intrinsically narrow, with a very low equivalent width compared with the other sources
\citep[just few tens of eV; see][]{Koliopanos2017},
so the lack of detection is not surprising.
A more in-depth study of the connections between the iron line intensity and
the determination of the drop in amplitude in the PF spectra are deferred to a future work.

In addition to  the energy-dependent Fourier amplitudes and its integral quantity, the PF,
other tools are  available from the study of the pulse profiles that rely on how the
peaks move in the phase space. The correlation and lag spectra offer an interesting
window to spot sudden changes occurring at the feature energy.
Changes in the lag spectrum are theoretically expected \citep{Schonherr2014},
and the observations of Her~X-1 and Cep~X-4 show a lag spectrum with a clear
drop in the lag value at energies close to \Ecyc; however, the observation
of 4U~1626$-$67 indicates an increase in the lag value, whereas no notable
change with respect to the overall trend is apparent in Cen~X-3.
Investigating the physical reasons of such results in the context of this small sample of
observations is beyond the scope of the present paper;  it will be addressed in a
future work compiling results of all cyclotron line sources detected by \nustar.
Similarly, the correlation spectra are worth a dedicate study; although no
clear change in the correlation values are spotted at the line positions,
the overall trends contain remarkable information, likely connected with
continuum variations (e.g., the correlation values are much more
constrained in a narrow range at 10--15\% of the maximum value, with the
notable exception  of the 4U~1626$-$67 observation,
which shows fast and steep decline  toward values of $\sim$\,0.4 at the energy
extremes).

Finally, the energy-resolved pulse color-maps offer a rapid and immediate
visualization of the peaks' multiplicity and their strength, which provides
a rapid inspection of the actual pulse behavior.

\section{Conclusions}

We  presented a quantitative approach to determining the
presence and characteristics of spectral features in the pulse-profile spectra
of XRPs. We applied our methodology to a small sample of sources to test
its efficacy in comparison to spectral modeling results.
Our findings offer strong indications about the consistency and robustness
of the method, though a much larger sample of observations, spanning different accretion conditions, need to
be explored to conclude whether this method can be straightforwardly used to
assess the presence and shape of cyclotron features in other sources,
where spectral detection is either lacking or poorly determined.
At the moment, this method could be used as a practical toolbox to
accompany the spectral analysis, but a clear understanding of the physical mechanisms that
produce the PF energy dependence and in general the pulse shape
would likely raise it to an independent probe of
physical measurement of XBPs spectral features.
We aim to extend this study to the full sample of cyclotron line sources observed by \nustar\
to evidence characteristic patterns.
We then plan to provide the elaborated data sets together with the methods employed in this paper
on a dedicated service where the various energy phases matrices can be re-used.\footnote{See the current sample at \href{https://renkulab.io/projects/carlo.ferrigno/ppanda-light/sessions/new?autostart=1}{this URL}.}

\begin{acknowledgements}
The research leading to these results has received funding
from the European Union’s Horizon 2020 Programme
under the AHEAD2020 project (grant agreement n. 871158).
EA, AD acknowledge funding from the Italian Space Agency, contract ASI/INAF n. I/004/11/4.
We made use of Heasoft and NASA archives for the \nustar data.
We developed our own timing code for the epoch folding, orbital correction,
building of time-phase and energy-phase matrices.
This code is based partly on available Python packages such as:
\texttt{astropy} \citep{astropy:2013,astropy:2018,astropy:2022},
\texttt{lmfit} \citep{lmfit},
\texttt{matplotlib} \citep{matplotlib},
\texttt{emcee} \citep{emcee},
\texttt{stingray} \citep{stingray},
\texttt{corner} \citep{corner},
\texttt{scipy} \citep{scipy}.
An online service that reproduces our current results is available on the Renku-lab platform of the Swiss Science Data Centre at \href{https://renkulab.io/projects/carlo.ferrigno/ppanda-light/sessions/new?autostart=1}{this link}.
We are grateful to Dr. P. Kretschmar and Dr. G. Cusumano for precious suggestions on this manuscript, and to the
language editor for a careful correction. All remaining issues are our responsibility.
\end{acknowledgements}

\bibliographystyle{aa}
\bibliography{refs,sw}

\appendix

\section{Pulsed fraction}\label{app:pf_def}
We briefly review and discuss here the most commonly used definitions of pulsed fraction (PF) present in the literature.

\begin{figure}
    \centering
    \includegraphics[width=\columnwidth]{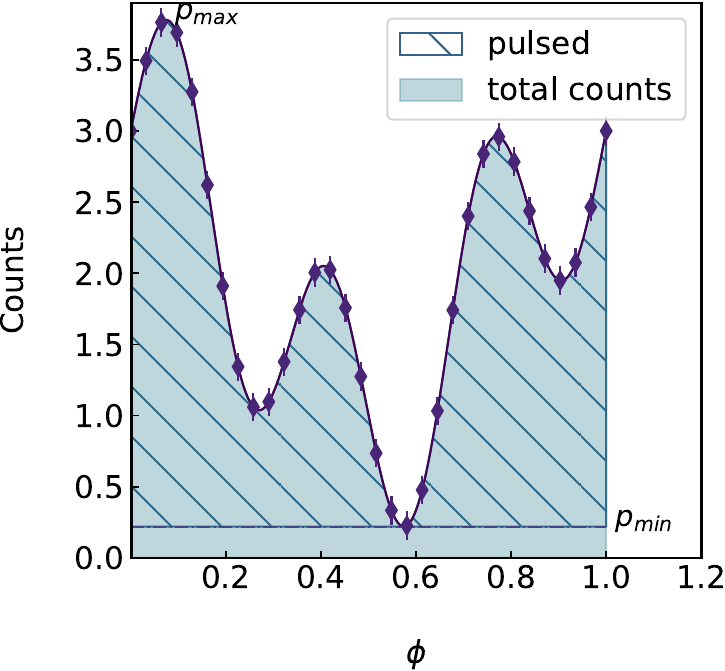}
    \caption{Generic pulse profile.}
    \label{fig:genericpulse}
\end{figure}

Because of its simple and straightforward calculation, one of the most commonly used definition of PF is the so-called PF$_{\mathrm{minmax}}$, defined as
\begin{equation}\label{eq:minmax}
\centering
\mathrm{PF}_{\mathrm{minmax}} = \frac{p_\mathrm{max}-p_\mathrm{min}}{p_\mathrm{max}+p_\mathrm{min}}
,\end{equation}
\noindent where $p_\mathrm{min}$ and $p_\mathrm{max}$ are respectively the minimum and the maximum values of the rate in the phase bins array
(as shown in Fig. \ref{fig:genericpulse}). The
PF$_{\mathrm{minmax}}$ is strongly biased on how much these values might change with the number
of phase bins of the profile, and because the computation is based only on two single reference values of the PF,
it is not sensitive to the full integral profile shape.

Another operational definition of PF is called area PF (PF$_{\mathrm{area}}$)
\begin{equation}\label{eq:pf_area}
\mathrm{PF}_{\mathrm{area}} = \frac{\Sigma_{i=0}^{N} (p_i - p_\mathrm{min})}{\Sigma_{i=0}^{N} p_i}
,\end{equation}
\noindent where $N$ is the number of phase bins and $p_\mathrm{min}$ is the minimum of the counts in the profile.
This definition directly represents the fraction of pulsed photons with respect to the total emission,
though it reduces the complexity of a profile to only one scalar value. As shown in \citet{An2015}, this
definition is in general affected by a larger uncertainty and potential bias when the minimum of the PF is
to be determined in a noisy profile at low statistics.

Finally, the root mean square PF (PF$_{\mathrm{rms}}$) measures the deviation of the pulse from its mean value.
The power of this method resides in the possibility to consider the complexity even of noisy pulse profiles. The
PF$_{\mathrm{rms}}$ can be obtained directly from the root mean square of the array of the $p_i$ values, and we  call this method
PF$_{\mathrm{rms,c}}$, and  through the power of Fourier coefficients.
It can be obtained as  \citep{Dhillon2009}
\begin{equation}\label{eq:pf_rms_c}
PF_{\mathrm{rms,c}} = \frac{\sqrt{ \Sigma_{i=0}^N \left[ (p_i - \bar p)^2 - \sigma_{p_i}^2 \right] / N}}{\bar p}
,\end{equation}
\noindent where $\bar p$ is the average count rate, $p_i$ the pulse profile, and $\sigma_{p_i}$ its uncertainty.

We analyzed two methods for the determination of PF through Fourier decomposition.
The most widely used \citep[see][and references therein]{An2015} is corrected for the intrinsic variance of the profile,
which we call PF$_{\mathrm{FFT,var}}$

\begin{equation}\label{eq:pf_rms}
\mathrm{PF}_{\mathrm{FFT,var}} =  \frac{\sqrt{2\sum_{k=1}^{ N_{\mathrm{harm}}}{((a_{k}^{2}+b_{k}^{2})-(\sigma_{a_{k}}^{2}+\sigma_{b_{k}}^{2}))}}}{a_{0}}
,\end{equation}
\noindent where N$_{\mathrm{harm}}$ is the number of harmonics used to describe the pulse profile, $a_{k}$ and $b_{k}$ are the Fourier coefficients,
and $\sigma_{a_{k}}$ and $\sigma_{b_{k}}$ their variances.\footnote{$a_{k} = \frac{1}{N} \sum_{i=1}^{N} c_{i} \cos \big( \frac{2 \pi k i}{N} \big)$, $b_{k} = \frac{1}{N} \sum_{i=1}^{N} c_{i} \sin \big( \frac{2 \pi k i}{N} \big)$ \\ $\sigma_{a_{k}}^{2}+\sigma_{b_{k}}^{2} =  \frac{1}{N^{2}} \sum_{i=1}^{N} \sigma_{c_{i}}^{2} \cos^{2} \big( \frac{2 \pi k i}{N} \big) + \frac{1}{N^{2}} \sum_{i=1}^{N} \sigma_{c_{i}}^{2} \sin^{2} \big( \frac{2 \pi k i}{N} \big)$}

Alternatively, we measured the rms variability through the direct computation of the Fourier decomposition using an FFT algorithm
(see Eq. (\ref{eq:fft}) in Sect.~\ref{sec:enphase-pf}).

To avoid including white noise, we limited ourselves to a number of harmonics $N_\mathrm{harm}$ so that the profile is
described at better than 90\% confidence level using the goodness of fit derived for Poissonian statistics by
\citet{Kaastra2017} (see Sect.~\ref{sec:enphase-pf}).

\begin{figure*}
    \centering
    \includegraphics[width = \columnwidth]{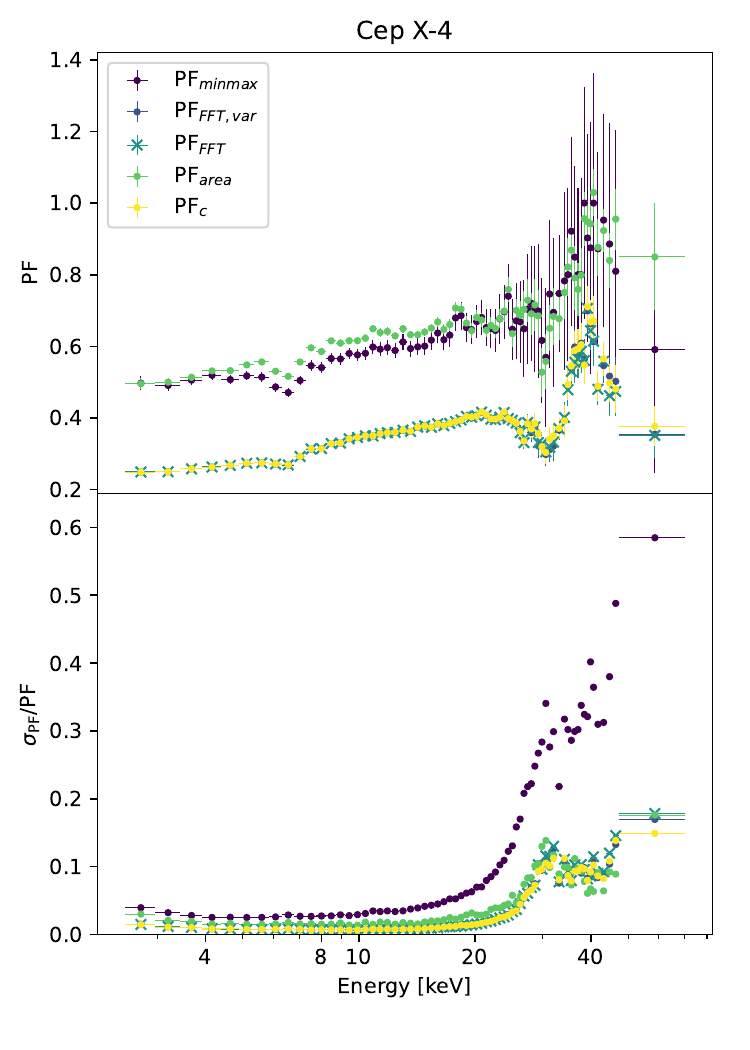}
    \includegraphics[width = \columnwidth]{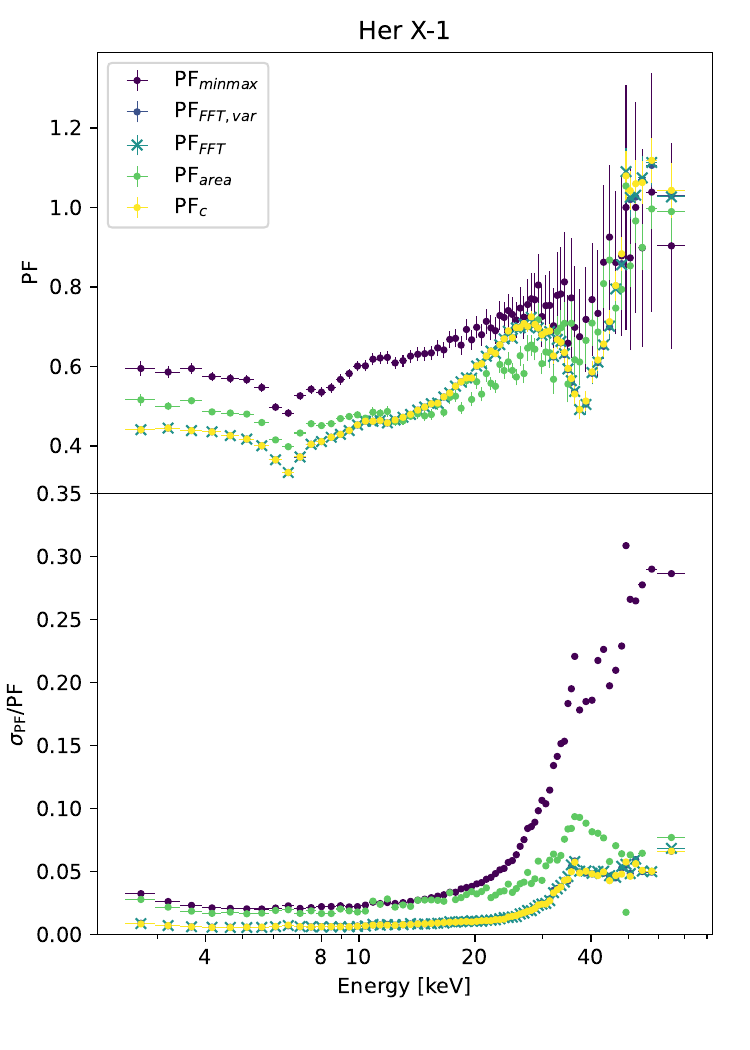}
    \caption{Comparison of   different methods used to obtain the pulsed fraction for \cepx  upper left) and Her X-1 ( upper right). In the lower panels the spectra of corresponding relative errors are shown. }\label{fig:PFmethodscomparison}
\end{figure*}

Figure~\ref{fig:PFmethodscomparison} shows the spectrum of the PF for \cepx (left) and Her X$\--$1 (right)
computed for the different definitions given above (Eq.~\ref{eq:minmax} - \ref{eq:fft}).
All the rms-like methods are equivalent, while the PF$_{\mathrm{minmax}}$ and the PF$_{\mathrm{area}}$ suffer
from uncertainties and bias, especially at higher energies where the statistics of the profiles are poorer.

Despite the global scaling factor among the different methods, rms methods appear more sensitive in describing the overall trend of the PF,
and do not severely suffer from poor statistics biases, as shown in the lower panels of Fig. \ref{fig:PFmethodscomparison},
where we show the relative uncertainty ($\sigma _{\mathrm{PF}}$/PF) for each method.
The rms-like methods, therefore, better suit the purpose of understanding the PF energy spectra that show local structures
associated with those obtained with spectral analysis.
We can choose any of the RMS methods, as the difference is well within the statistical uncertainties and never exceed 1\%.
The equivalence of methods implies
that the correction or the variance is negligible, at least once the pulse profiles have the minimum S/N that we adopt.

\begin{figure}
    \centering
    \includegraphics[width=\columnwidth]{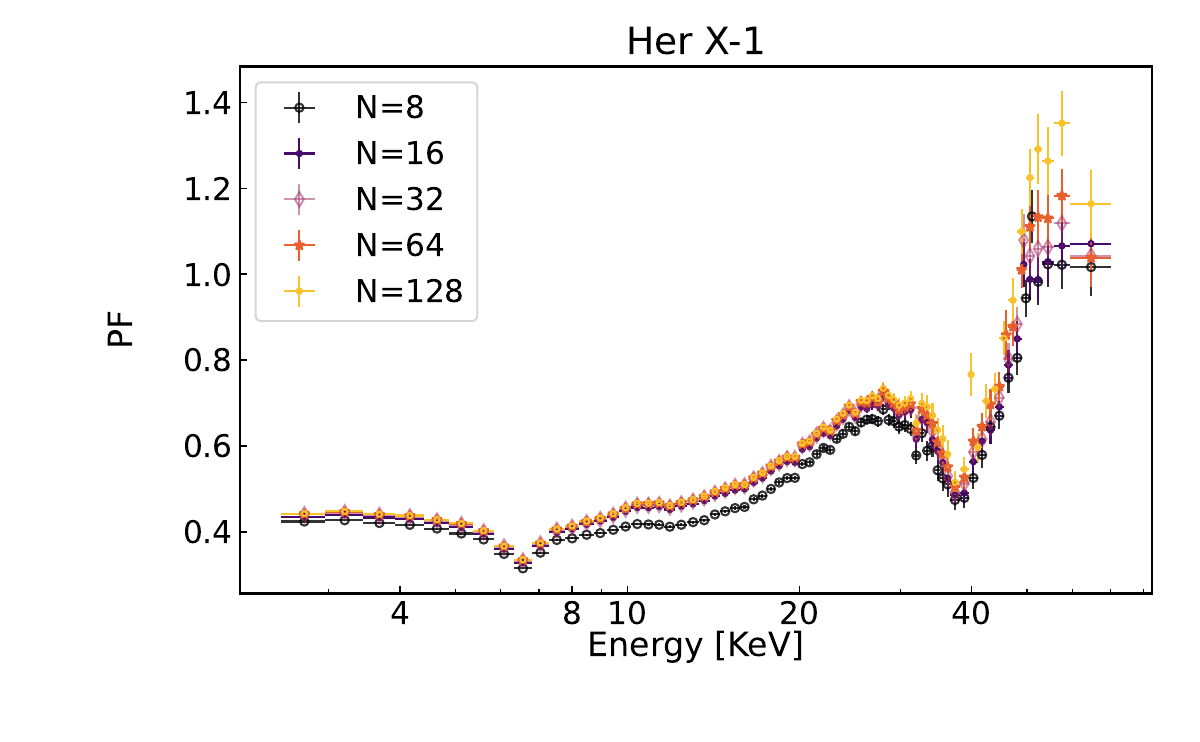}
    \caption{PF$_{\mathrm{FFT}}$ obtained with different N$_{\mathrm{bins}}$ for the pulse profile of Her X $\--$1 analyzed in this work.}
    \label{fig:her1_nbins_fft}
\end{figure}
\begin{figure}
    \centering
    \includegraphics[width=0.9\columnwidth]{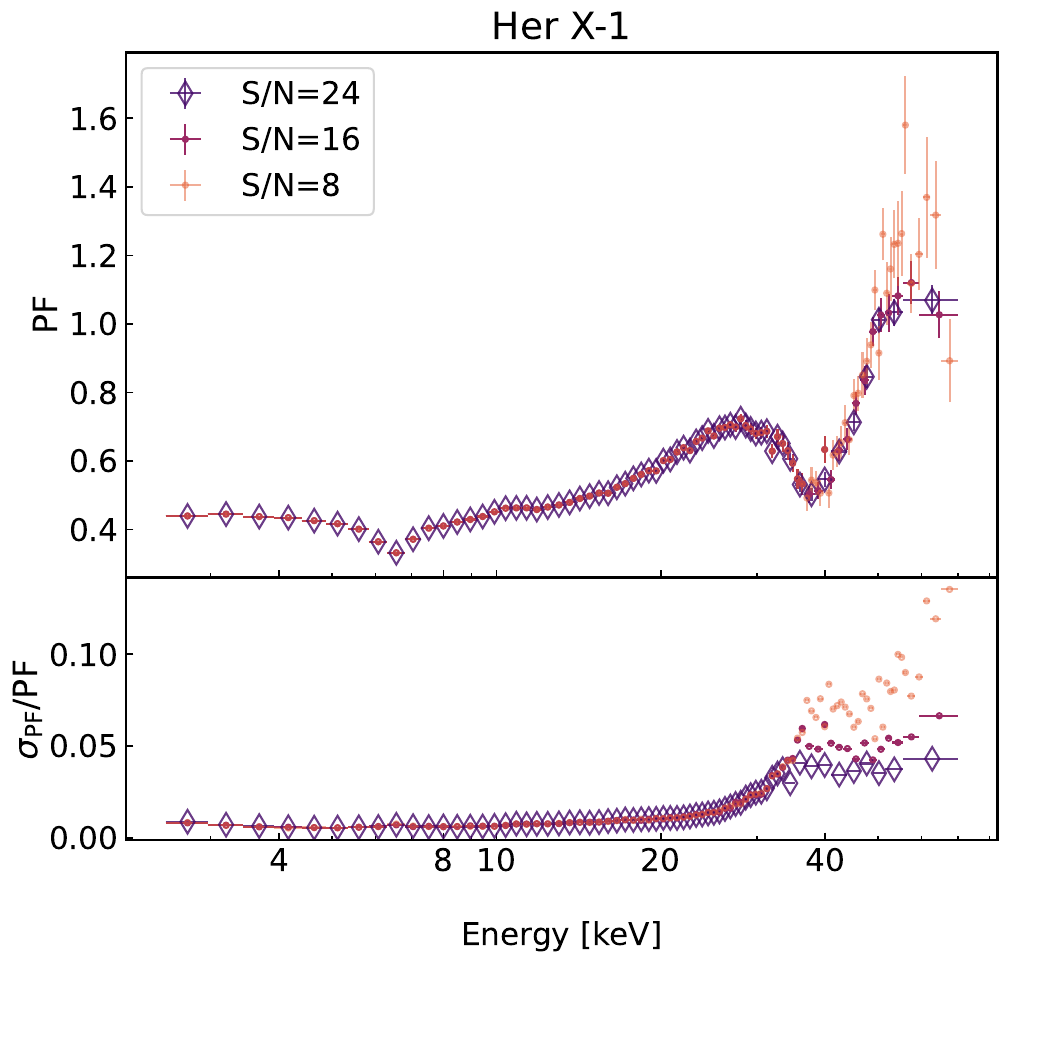}
    \caption{PF$_{\mathrm{FFT}}$ obtained with different S/N values for the pulse profile of Her X $\--$1 analyzed in this work. N$_{\mathrm{bins}}$ = 32.}
    \label{fig:her_snr}
\end{figure}

These methods are robust against the number of chosen phase bins if in the range 16 $\leq $ N$_{\mathrm{bins}}  \leq $ 64 (see Fig. \ref{fig:her1_nbins_fft}).
A low value of N$_{\mathrm{bins}}$ underestimates the PF at low energies where the profiles are more complex.
Finally, it should be also  noted that the choice of the minimum S/N has an impact on the fitting of the PF spectrum,
as can be seen in Fig.~\ref{fig:her_snr}.

\section{EMCEE corner plots and fit results\label{app:emcee_plots}}

In this Appendix we show the corner plots
for the best-fitting values and associated errors
of the PF spectra fits presented in Tables~\ref{tab:4u1626_cyc}, \ref{tab:herx1_harmonics},
\ref{tab:cen_x-3_line_parameters}, and \ref{tab:cep_x-4_line_parameters}.

\begin{figure}
    \centering
    \begin{tabular}{c}
        \includegraphics[width=\columnwidth]{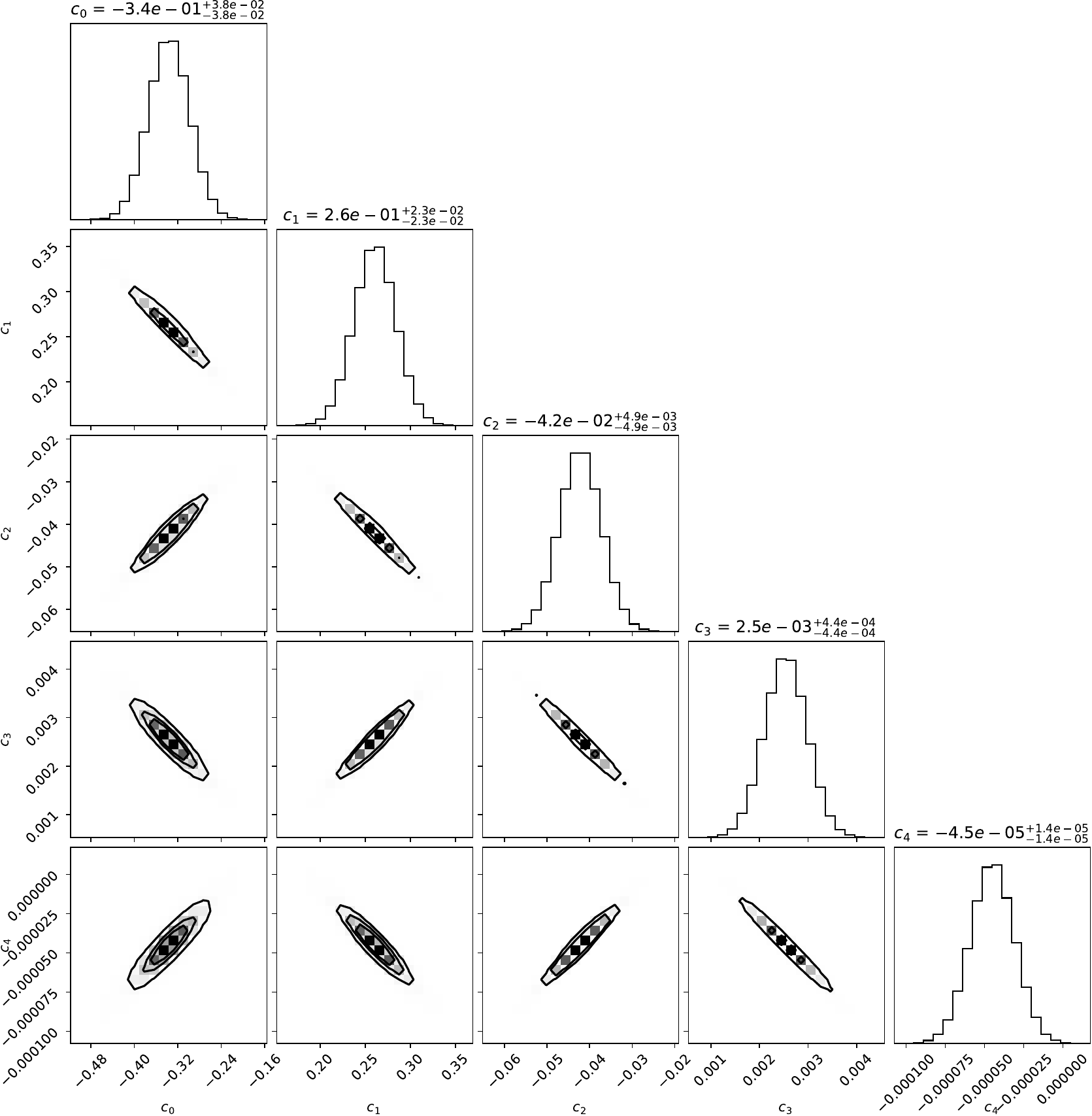}     \\
        \includegraphics[width=\columnwidth]{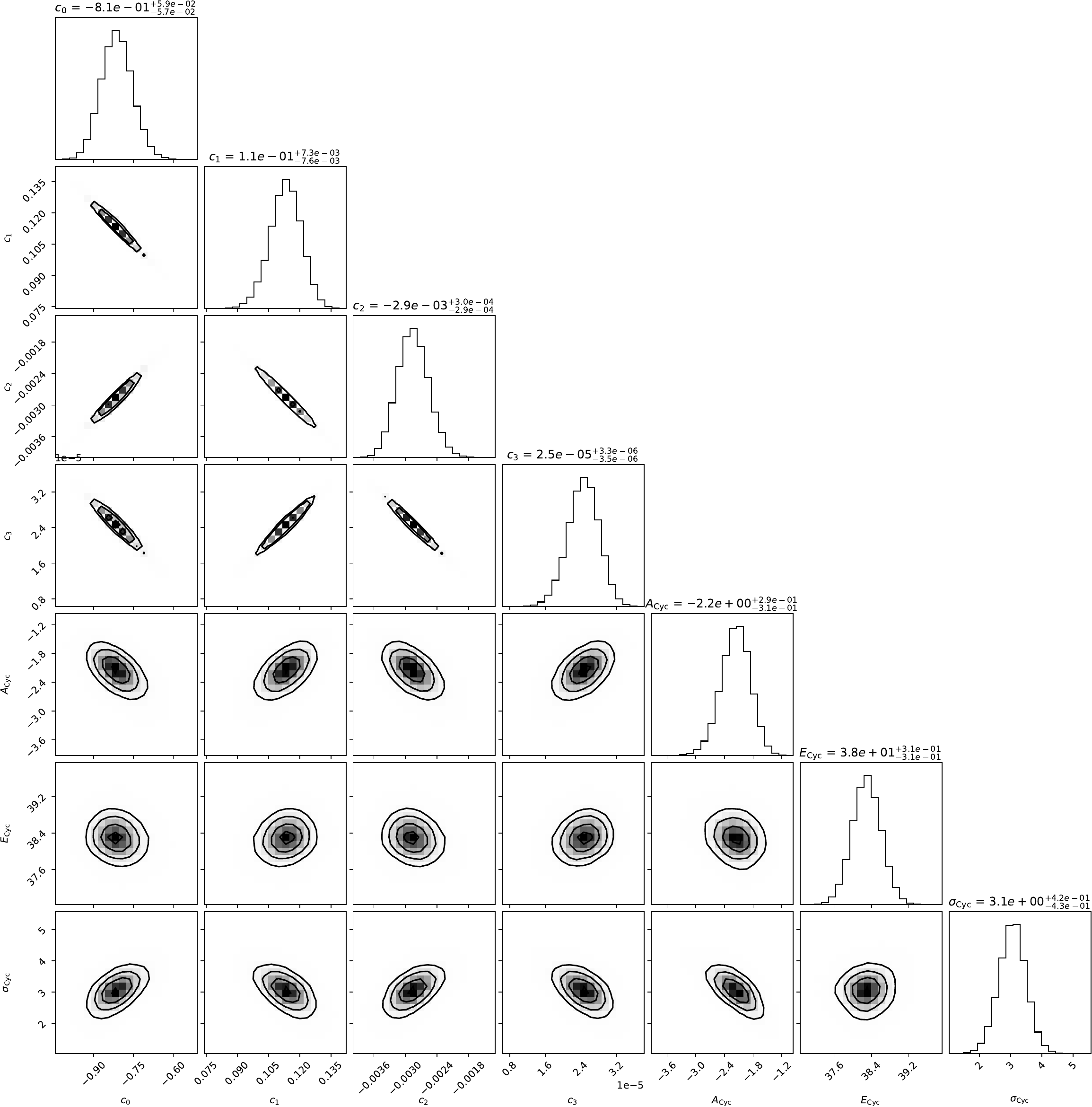}     \\
    \end{tabular}
    \caption{Corner plots for the 4U~1626$-$67 PF spectrum fit. The upper and lower panels show the
        soft and hard band fits, respectively.}
    \label{fig:4u162emcee}
\end{figure}
\newpage

\begin{figure}
    \centering
    \begin{tabular}{c}
        \includegraphics[width=\columnwidth]{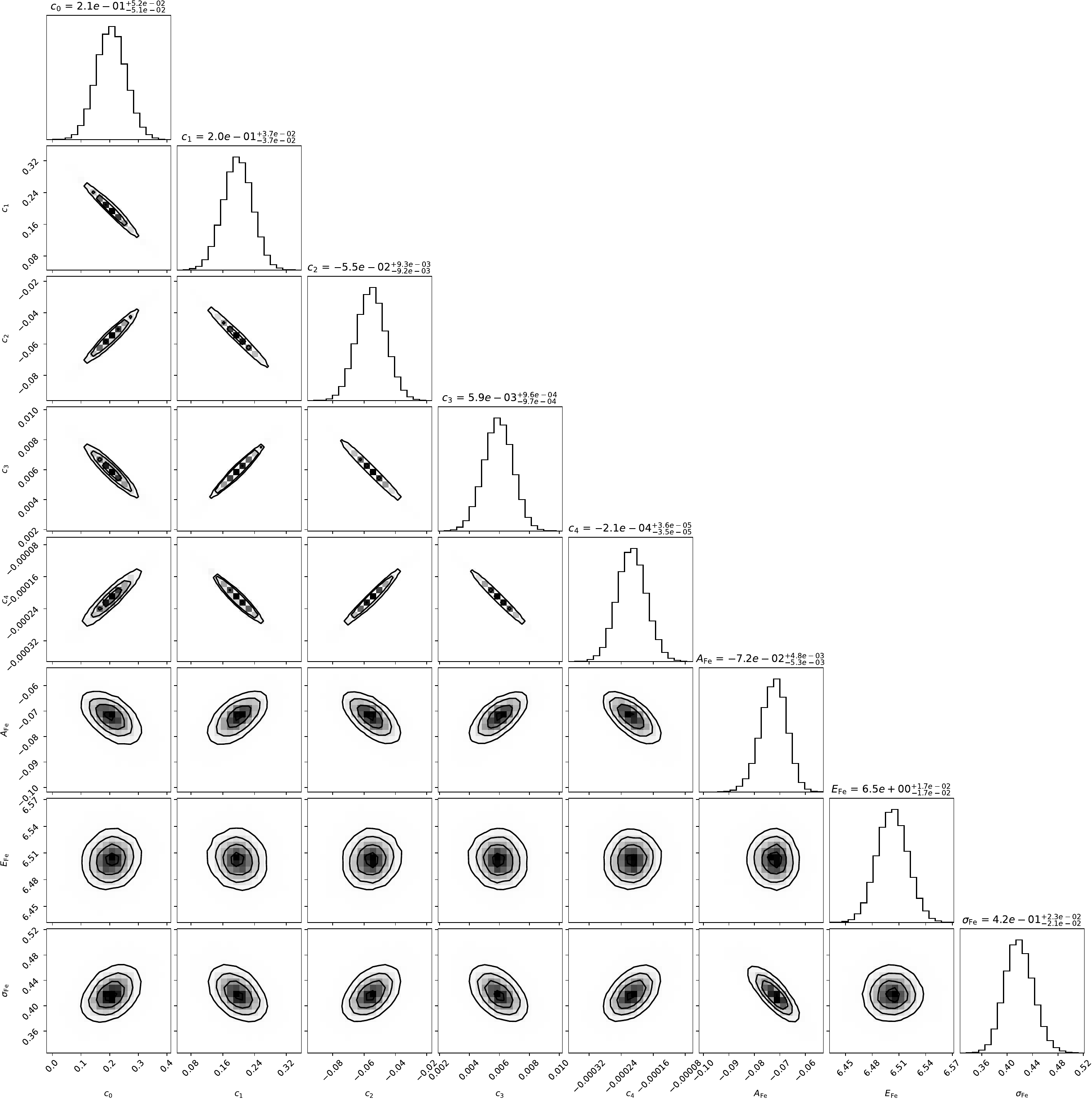}     \\
        \includegraphics[width=\columnwidth]{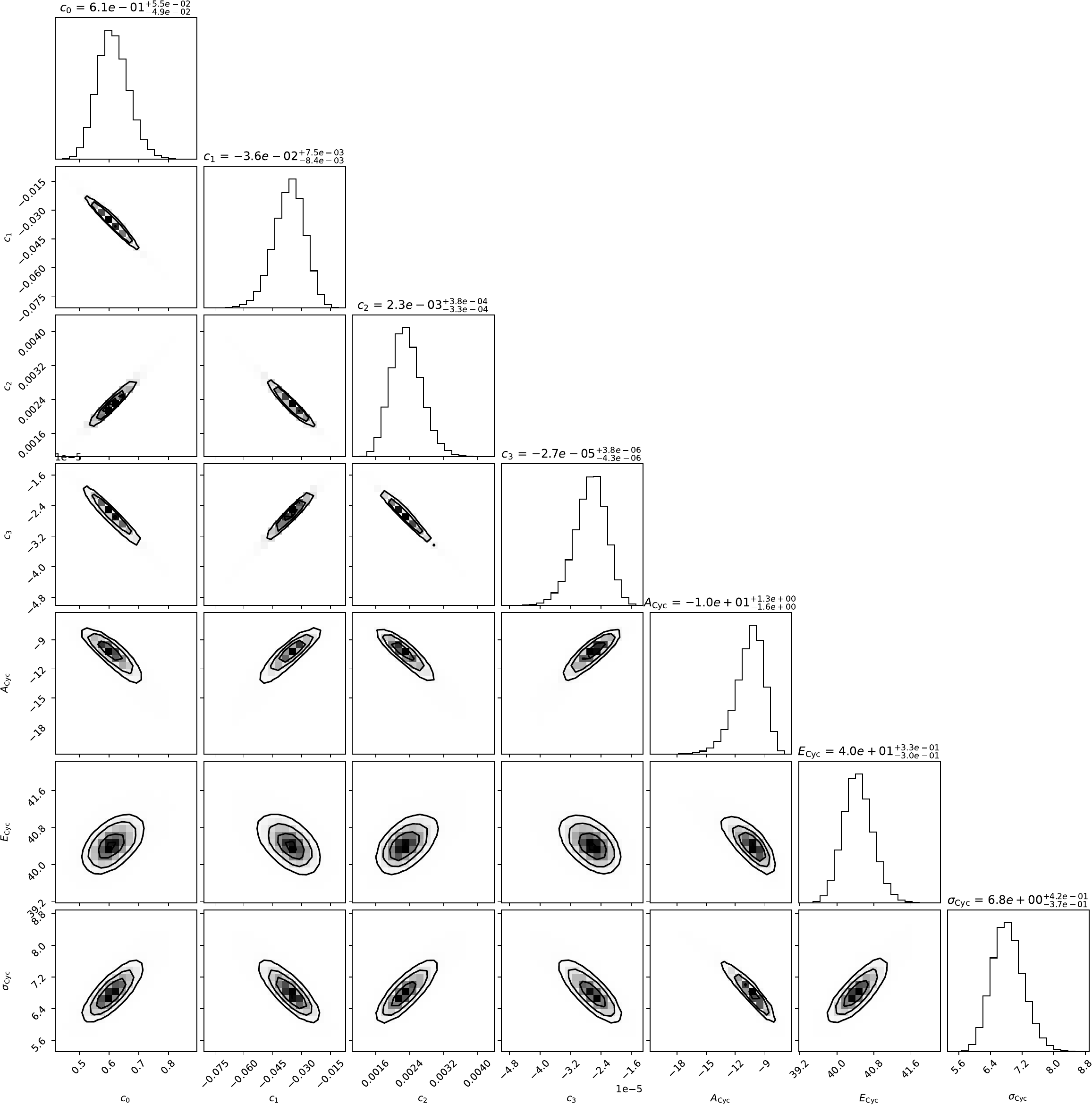}     \\
    \end{tabular}
    \caption{Corner plots for the Her X-1 pulsed fraction spectrum fit. The upper and lower panels show the
        soft and hard band fits, respectively.}
    \label{fig:herx1emcee}
\end{figure}

\begin{figure}
    \centering
    \begin{tabular}{c}
        \includegraphics[width=\columnwidth]{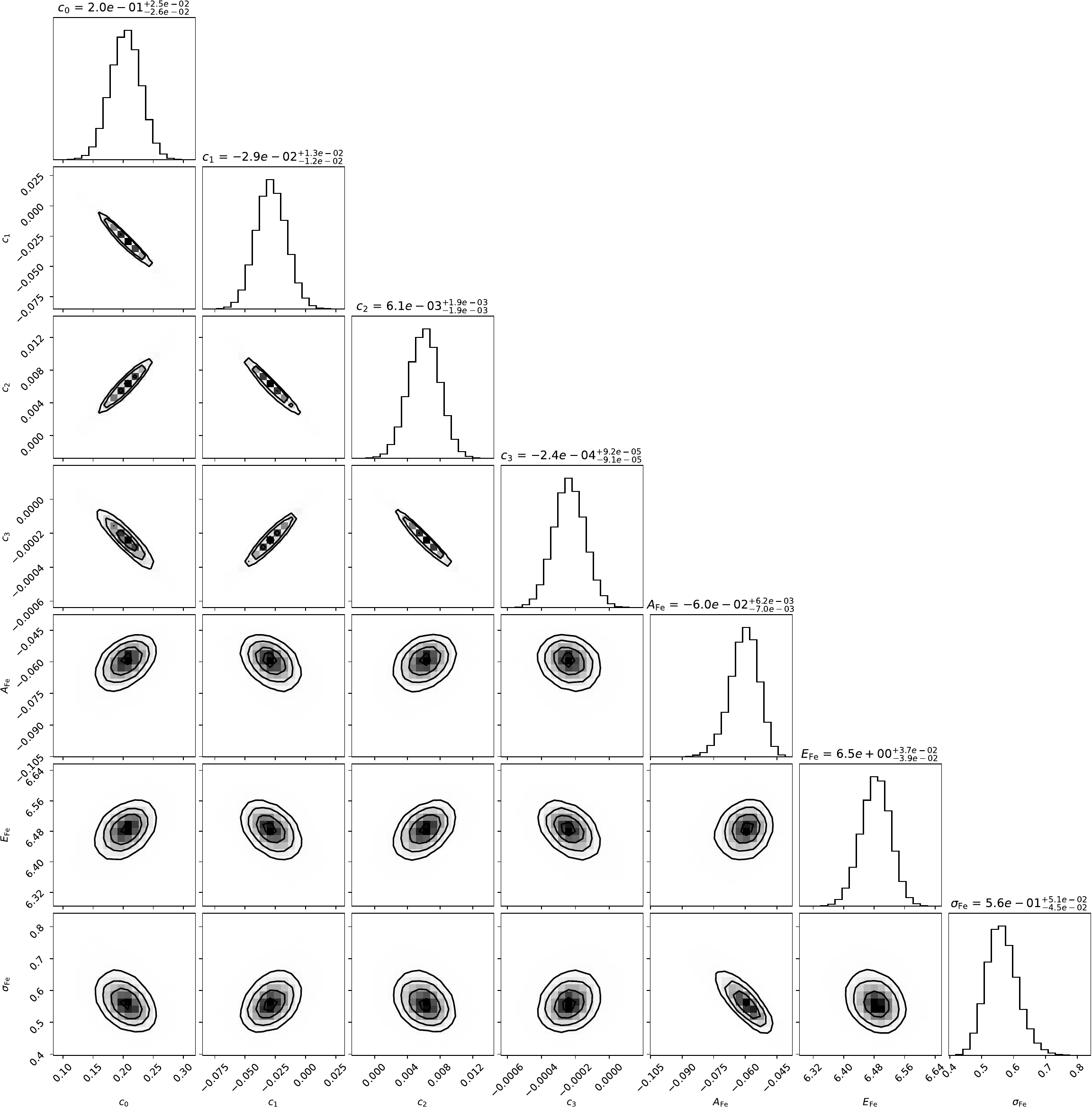}     \\
        \includegraphics[width=\columnwidth]{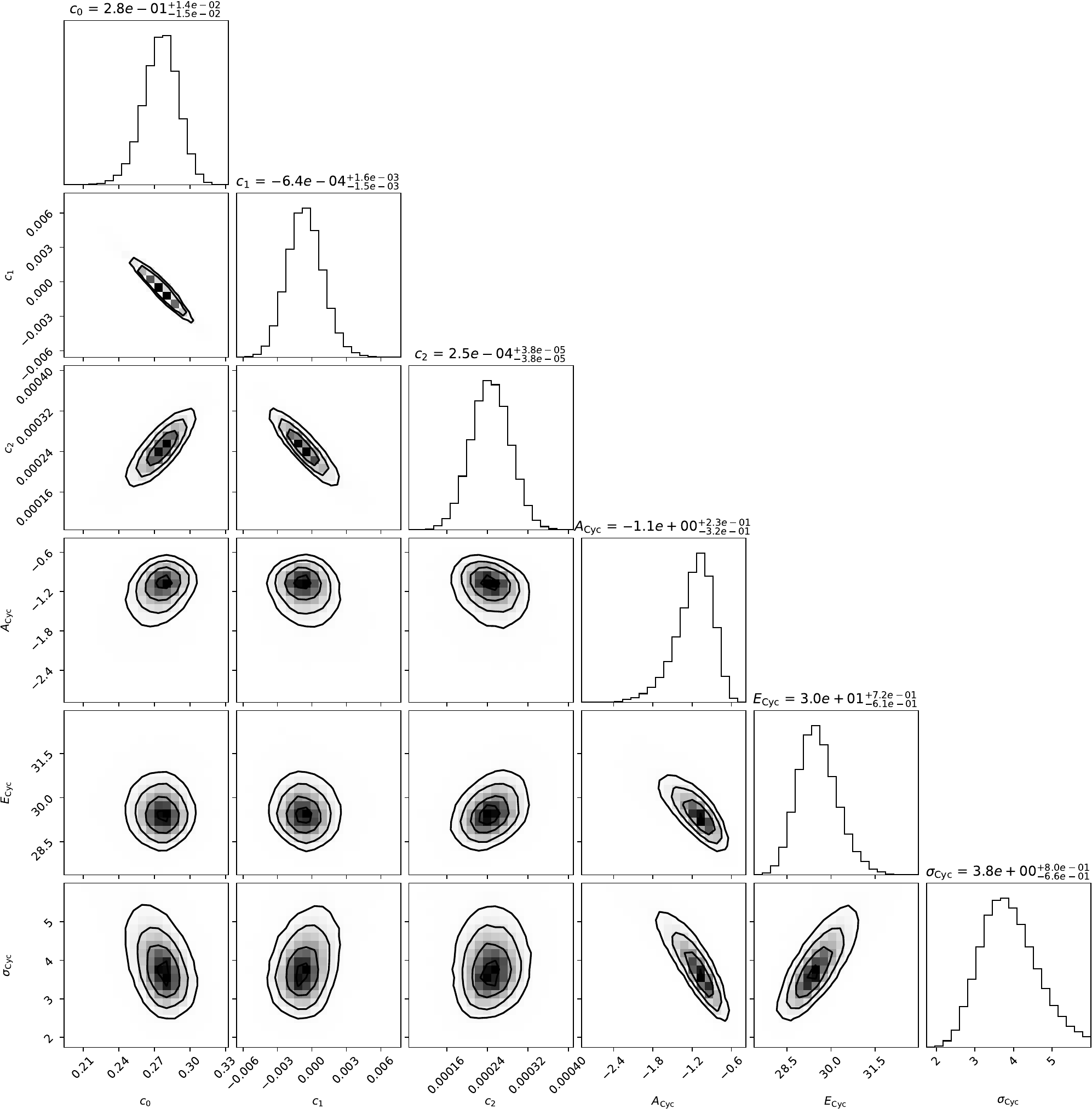}     \\
    \end{tabular}
    \caption{Corner plots for the Cen~X-3 pulsed fraction spectrum fit. The upper and lower panels show the
        soft and hard band fits, respectively.}
    \label{fig:cenx3emcee}
\end{figure}

\begin{figure}
    \centering
    \begin{tabular}{c}
        \includegraphics[width=\columnwidth]{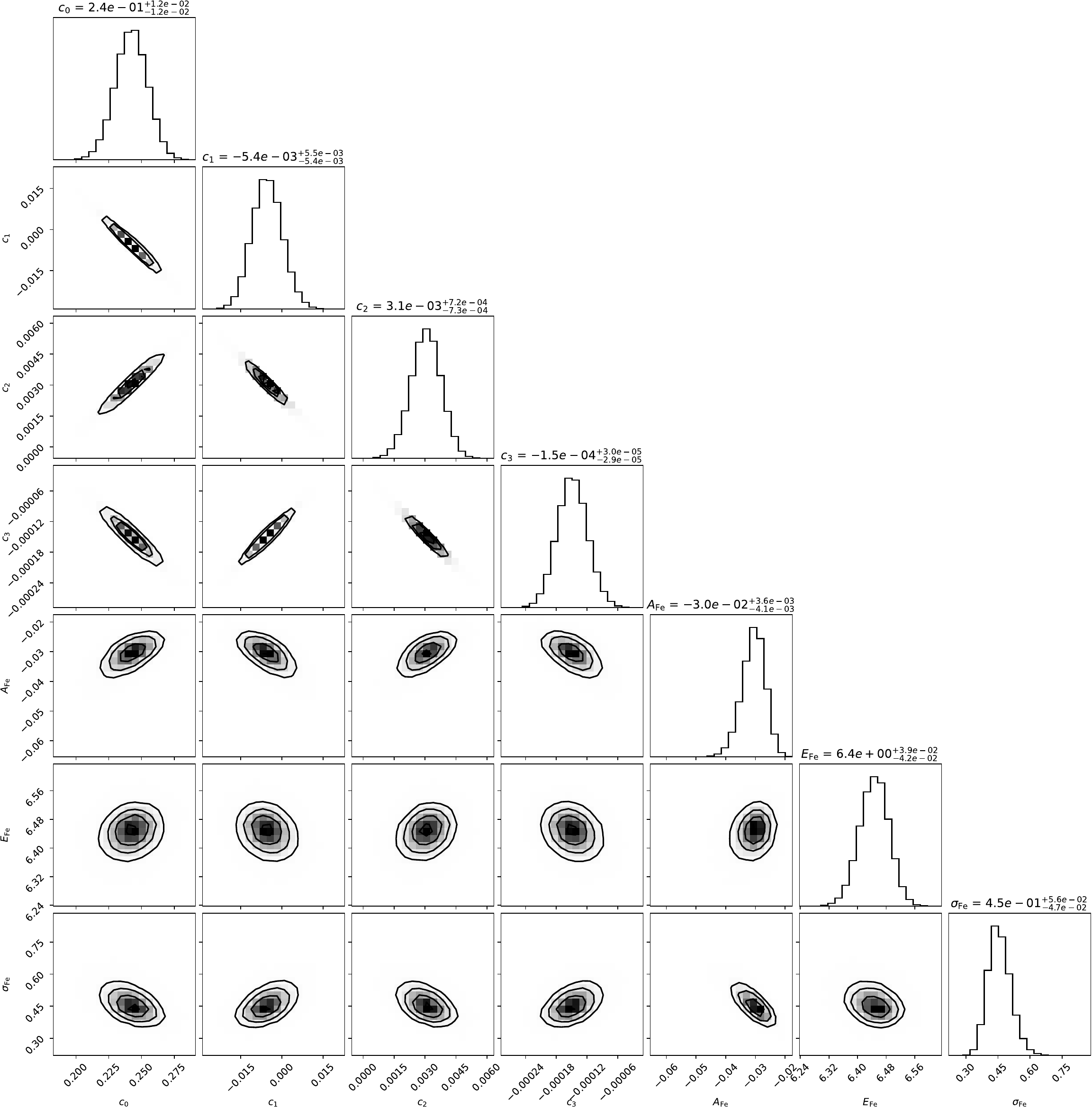}     \\
        \includegraphics[width=\columnwidth]{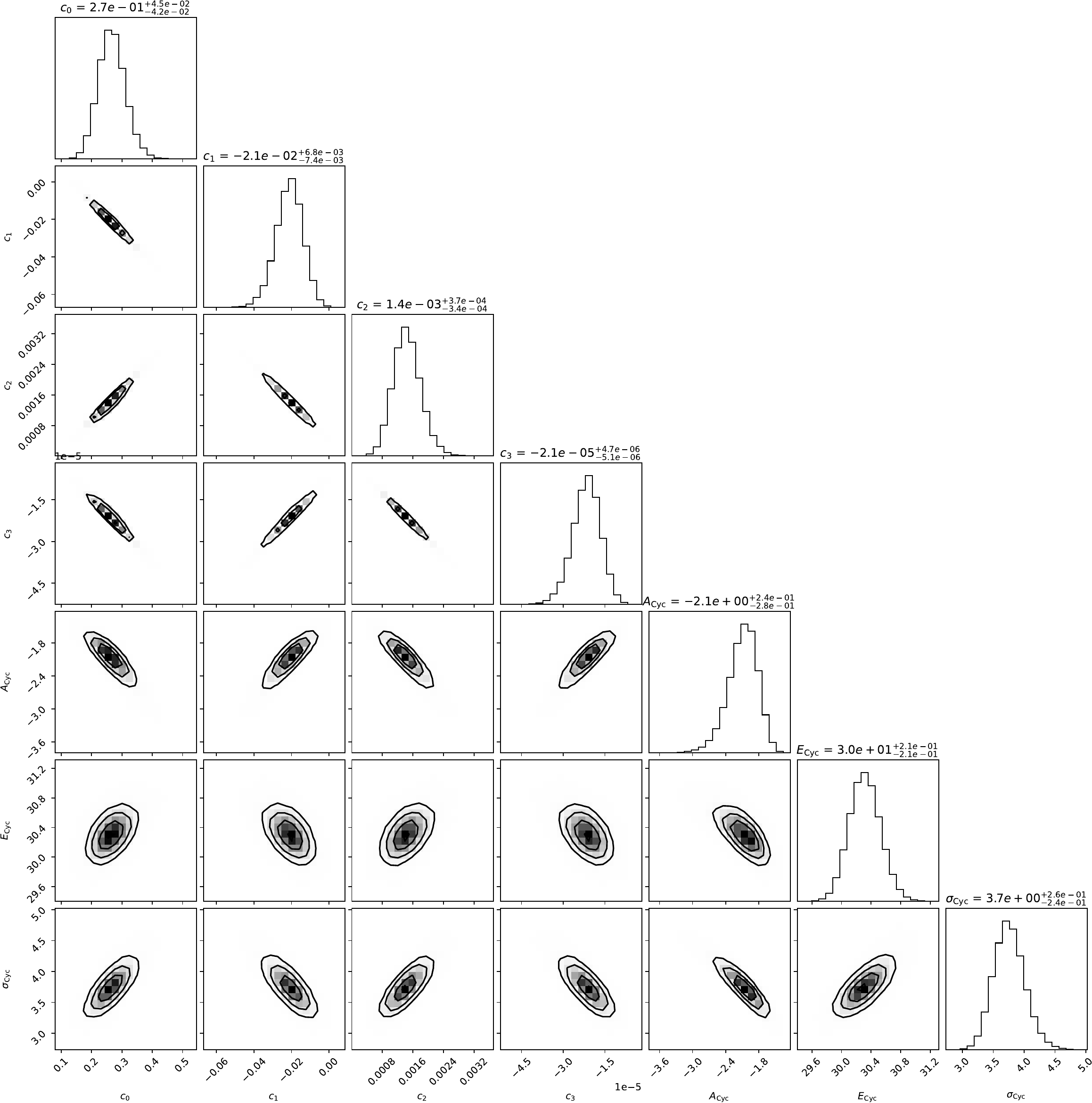}     \\
    \end{tabular}
    \caption{Corner plots for the \cepx pulsed fraction spectrum fit. The upper and lower panels show the
        soft and hard band fits, respectively.}
    \label{fig:cepx4emcee}
\end{figure}

\end{document}